\documentclass[preprint, 3p, table, xcdraw, authoryear]{elsarticle}

\usepackage{lineno}
\modulolinenumbers[1]

\usepackage{tikz}
\NewDocumentCommand{\rot}{O{90} O{1em} m}{\makebox[#2][l]{\rotatebox{#1}{#3}}}%

\usepackage{setspace}
\journal{Journal of \LaTeX\ Templates}
\makeatletter
\def\ps@pprintTitle{%
    \let\@oddhead\@empty
    \let\@evenhead\@empty
    \def\@oddfoot{\footnotesize\itshape
         {Preprint} \hfill\today}%
    \let\@evenfoot\@oddfoot
    }
\makeatother

\usepackage{doi}
\usepackage{ulem}
\useunder{\uline}{\ul}{}
\usepackage{amssymb}
\usepackage{amsmath}
\usepackage[nameinlink]{cleveref}
\usepackage{caption}
\usepackage{subcaption}
\usepackage[export]{adjustbox}
\usepackage{bm}
\usepackage{bbm}
\usepackage{multicol}

\usepackage{booktabs}
\usepackage{multirow}
\usepackage[flushleft]{threeparttable}

\usepackage{hyperref}
\hypersetup{
  colorlinks   = true, %Colours links instead of ugly boxes
  urlcolor     = blue, %Colour for external hyperlinks
  linkcolor    = blue, %Colour of internal links
  citecolor   = blue %Colour of citations
}
\newcommand{\beginsupplement}{%
        \setcounter{table}{0}
        \renewcommand{\thetable}{S\arabic{table}}%
        \setcounter{figure}{0}
        \renewcommand{\thefigure}{S\arabic{figure}}%
        \renewcommand{\thesubsection}{\Alph{subsection}}
     }
\begin{document}

%TC:ignore
\begin{frontmatter}

    \title{Evaluating geospatial context information for travel mode detection}

    \author[ikg]{Ye Hong\corref{correspondingauthor}}
    \cortext[correspondingauthor]{Corresponding author}
    \ead{hongy@ethz.ch}

    \author[ikg]{Emanuel Stüdeli}
    \ead{estuedeli@student.ethz.ch}

    \author[ikg]{Martin Raubal}
    \ead{mraubal@ethz.ch}

    \address[ikg]{Institute of Cartography and Geoinformation, ETH Zurich}

    \begin{abstract}
        Detecting travel modes from global navigation satellite system (GNSS) trajectories is essential for understanding individual travel behavior and a prerequisite for achieving sustainable transport systems.
        While studies have acknowledged the benefits of incorporating geospatial context information into travel mode detection models, few have summarized context modeling approaches and analyzed the significance of these context features, hindering the development of an efficient model.
        Here, we identify context representations from related work and propose an analytical pipeline to assess the contribution of geospatial context information for travel mode detection based on a random forest model and the SHapley Additive exPlanation (SHAP) method.
        Through experiments on a large-scale GNSS tracking dataset, we report that features describing relationships with infrastructure networks, such as the distance to the railway or road network, significantly contribute to the model's prediction.
        Moreover, features related to the geospatial point entities help identify public transport travel, but most land-use and land-cover features barely contribute to the task.
        We finally reveal that geospatial contexts have distinct contributions in identifying different travel modes, providing insights into selecting appropriate context information and modeling approaches.
        The results from this study enhance our understanding of the relationship between movement and geospatial context and guide the implementation of effective and efficient transport mode detection models.
    \end{abstract}

\end{frontmatter}

% \linenumbers

% \tableofcontents
%TC:endignore
\section{Introduction}\label{sec:intro}

Knowledge regarding individuals' usage of travel modes is an indispensable element in contemporary travel behavior studies.
Travel mode choices are formed due to the everyday needs and constraints of individuals~\citep{hagerstrand_what_1970} and are generally influenced by travel-related factors such as cost, time, accessibility and comfort~\citep{ortuzar_modelling_2011}. As a result, it is not uncommon for mobility systems to be evaluated and compared based on the current modal splits of the overall population~\citep{lee_what_2022} or their experienced modal shifts~\citep{buehler_reducing_2017} for reflecting the travel behavior situations in a defined area.
Besides, individuals' choices in travel modes reflect personal travel preferences and habits~\citep{hong_conserved_2023}, whose in-depth understanding benefits traffic modelling~\citep{horni_multi_agent_2016} and transport planning~\citep{molloy_mobis_2022}.
Studies on individual travel mode choices have been particularly relevant nowadays due to the growing impact of mobility on the environment~\citep{erhardt_transportation_2019}. Knowing travel mode shares for various activity-travel types is an essential prerequisite for estimating environmental impacts~\citep{bohm_gross_2022} and promoting new mobility concepts, such as mobility as a service (MaaS), to achieve a more efficient and sustainable transport system~\citep{reck_mode_2022}.

With the flourishing of information and communication technologies (ICT) \citep{bucher_location_2019}, the primary approach to collecting travel behavior information has gradually evolved from conventional travel surveys, where people are asked to complete questionnaires online, on paper, or by telephone, to sensors and devices that record location information automatically~\citep{stopher2008search, wang_applying_2018, raubal_geosmartness_2021}.
Smartphone applications that utilize built-in global navigation satellite system (GNSS) sensors stand out due to their high data quality, low implementation costs, and easy administration~\citep{shen_review_2014, marra_developing_2019}.
These smartphone-GNSS datasets contain location information recorded at a high spatial and temporal resolution, enabling continuous monitoring of individuals' whereabouts~\citep{barbosa_human_2018}.
Moreover, GNSS tracking data is suitable for uncovering the complex relationships between human mobility and its surrounding environments~\citep{rout_using_2021, Hong_context_2023}.
Despite these satisfactory properties, smartphone GNSS sensors do not support an automatic inference about high-level travel information, including on which travel mode the individual is currently conducting the travel.

Over the past decade, many studies have contributed to the automatic detection of travel mode, which is now regarded as a standard processing step for GNSS tracking-based travel behavior studies~\citep{shen_review_2014}.
Early attempts are dominated by rule-based heuristics or fuzzy logic methods, where domain experts design rules to differentiate between travel modes~\citep{chen_evaluating_2010, schuessler_processing_2009}.
Recent interests have been gradually switched to machine learning (ML) based methods that can effectively learn non-linear relationships directly from data~\citep{wang2018travel}, thus increasing the mode detection performance for large-scale real-world tracking datasets~\citep{wu_travel_2022}.
Unlike rule-based heuristics with well-designed decision boundaries, the interpretability of ML models remains low~\citep{prelipcean_transportation_2017, xiao_identifying_2017}. With little knowledge of the inner working mechanism, we cannot safely assess the models' ability to generalize to new populations and to transfer to new geographical areas.
Moreover, researchers have increasingly recognized the importance of geospatial context information, such as bus stops, land cover, and points of interests (POIs), in characterizing movements with different travel modes~\citep{semanjski_spatial_2017}.
By representing context information as features and incorporating them into the framework, detection performance can be significantly improved~\citep{roy2022assessing, zeng_trajectory_2023}.
However, the relationship between movement and context can be depicted in multiple ways. There is currently no consensus on which types of geospatial context information are most critical for mode detection.
As a result, detection approaches create a set of features based on their interpretation, typically including more contextual variables than necessary as input to ML models.

To address these gaps, we propose a pipeline to systematically evaluate the contribution of geospatial context information for travel mode detection.
Specifically, we thoroughly review recent literature to identify common and natural context representations, which are then implemented based on open-source geospatial data.
We develop a random forest (RF) model, reported as one of the best-performing ML models for this task.
Finally, we apply SHapley Additive exPlanation (SHAP), a type of feature attribution method, to the trained RF model to evaluate the features' impacts.
In short, our contributions are summarized as follows:
\begin{itemize}
    \item We implement a comprehensive set of features for travel mode detection. The features are summarized from previous research and describe motion characteristics and geospatial context information, covering a broad spectrum of context modeling approaches.
    \item We quantitatively evaluate the importance of geospatial context features to the detection performance. Our results suggest that geospatial network features contribute the most, and the separation of travel modes can benefit from features describing their respective infrastructures.
    \item We conduct experiments on a large-scale GNSS tracking dataset. The dataset includes individual movements within Switzerland and involves seven travel modes: bicycle, boat, bus, car, train, tram, and walk, which is one of the largest studies in terms of geographical areas and detection scheme.
    \item We include only open-source geospatial data in the feature construction, ensuring the framework's reproducibility and generalizability. We open-source our framework to provide a benchmark for further reference\footnote{The source code is available at \url{https://github.com/mie-lab/mode_detect}}.
\end{itemize}

\section{Related work}\label{sec:related}

Travel mode detection is the task of inferring the travel mode utilized by an individual given a movement trajectory.
Here, we focus on detecting travel modes from trajectories recorded using smartphone GNSS sensors, as they provide dense mobility traces that can lead to fine-grained mode detection results~\citep{huang2019transport}.
The general approach for this line of mode detection studies involves three steps~\citep{shen_review_2014, prelipcean_transportation_2017, zeng_trajectory_2023}.
First, rule-based algorithms are designed to detect mode transfer points (MTPs) from raw GNSS track points, which segment the continuous trajectory into stages conducted with a single travel mode.
Then, characteristic features such as speed or heading are extracted from the stages.
Finally, rule-based heuristics, statistical methods, or ML classifiers are developed to infer the travel mode based on these features.
As the detection of MTPs has become the de-facto preprocessing standard~\citep{Tsui2006enhanced, schuessler_processing_2009}, we discuss the feature extraction and the method development in the following section.

\subsection{Feature extraction and importance assessment}

To identify typical features for travel mode detection, we consult review papers~\citep{shen_review_2014, gong2014deriving, prelipcean_transportation_2017} and select representative studies published in recent years. An overview of features implemented in the reviewed literature can be found in Table~\ref{tab:lit}. We distinguish between features that characterize the movement (motion feature) and features that describe the relationship between motion and context information (geospatial context feature).

\addtolength{\tabcolsep}{-2pt}
\newcommand{\cm}{\checkmark}

\begin{table}[!ht]

    \centering
    \caption[Summary of input features in related literature.]{Summary of input features in related literature (PT: public transport). }
    \label{tab:lit}
    \begin{adjustbox}{width=1.0\textwidth,center=\textwidth}
        \begin{threeparttable}

            \begin{tabular}{c >{\columncolor[gray]{0.8}}cc>{\columncolor[gray]{0.8}}cc>{\columncolor[gray]{0.8}}cc>{\columncolor[gray]{0.8}}cc>{\columncolor[gray]{0.8}}c c>{\columncolor[gray]{0.8}}cc>{\columncolor[gray]{0.8}}cc>{\columncolor[gray]{0.8}}cc>{\columncolor[gray]{0.8}}cc>{\columncolor[gray]{0.8}}cc>{\columncolor[gray]{0.8}}cc>{\columncolor[gray]{0.8}}cc>{\columncolor[gray]{0.8}}cc}
                \toprule
                \multirow{2}{*}{}                    & \multicolumn{9}{c}{Motion feature} & \multicolumn{16}{c}{Geospatial context feature}                                                                                                                                                                                                                                                                                                                                                                                                                                                                                                             \\ \cmidrule(lr){2-10}\cmidrule(lr){11-26}
                                                     & \rot{Speed}                        & \rot{Acceleration}                              & \rot{Jerk} & \rot{Bearing rate} & \rot{Length} & \rot{Duration} & \rot{Altitude} & \rot{GNSS accuracy} & \rot{Trackpoint distance} & \rot{Road network} & \rot{PT route} & \rot{Railway network} & \rot{Tram network} & \rot{Subway network} & \rot{Cycling path} & \rot{Walking path} & \rot{Bus stop} & \rot{Subway station} & \rot{Train station} & \rot{PT timetable} & \rot{Real-time bus location} & \rot{Water body} & \rot{Public green space} & \rot{Commercial center} & \rot{Residential area} \\

                \midrule

                \citet{patterson2003inferring}       & \cm                                &                                                 &            &                    &              &                &                &                     &                           &                    & \cm            &                       &                    &                      &                    &                    & \cm            &                      &                     &                    &                              &                  &                          &                         &                        \\
                \citet{Tsui2006enhanced}             & \cm                                &                                                 &            &                    &              &                &                & \cm                 &                           &                    & \cm            &                       &                    &                      &                    &                    &                & \cm                  &                     &                    &                              &                  &                          &                         &                        \\
                \citet{zheng2008learning}            & \cm                                & \cm                                             &            &                    & \cm          &                &                &                     &                           &                    &                &                       &                    &                      &                    &                    &                &                      &                     &                    &                              &                  &                          &                         &                        \\
                \citet{stopher2008search}            & \cm                                &                                                 &            &                    & \cm          &                &                &                     &                           & \cm                & \cm            & \cm                   &                    &                      &                    &                    & \cm            &                      &                     &                    &                              &                  &                          &                         &                        \\
                \citet{chen_evaluating_2010}         & \cm                                &                                                 &            &                    &              & \cm            &                & \cm                 &                           &                    & \cm            &                       &                    &                      &                    &                    & \cm            &                      &                     &                    &                              &                  &                          &                         &                        \\
                \citet{stenneth2011transportation}   & \cm                                & \cm                                             &            & \cm                &              &                &                & \cm                 &                           &                    &                & \cm                   &                    &                      &                    &                    & \cm            &                      &                     &                    & \cm                          &                  &                          &                         &                        \\
                \citet{gong2012gps}                  & \cm                                & \cm                                             &            &                    &              & \cm            &                &                     &                           &                    &                &                       &                    &                      &                    &                    & \cm            & \cm                  & \cm                 &                    &                              &                  &                          &                         &                        \\
                \citet{biljecki2013transportation}   & \cm                                &                                                 &            &                    &              &                &                &                     &                           & \cm                & \cm            & \cm                   &                    &                      & \cm                & \cm                & \cm            & \cm                  & \cm                 &                    &                              & \cm              &                          &                         &                        \\
                \citet{xiao2015travel}               & \cm                                & \cm                                             &            &                    & \cm          &                &                &                     &                           &                    &                &                       &                    &                      &                    &                    &                &                      &                     &                    &                              &                  &                          &                         &                        \\
                \citet{rasmussen2015improved}        & \cm                                & \cm                                             &            &                    &              &                &                &                     &                           & \cm                &                & \cm                   &                    &                      &                    &                    & \cm            &                      &                     &                    &                              &                  &                          &                         &                        \\
                \citet{feng2016comparison}           & \cm                                & \cm                                             &            &                    & \cm          &                &                & \cm                 &                           & \cm                &                &                       & \cm                & \cm                  &                    &                    &                &                      &                     &                    &                              &                  &                          &                         &                        \\
                \citet{xiao_identifying_2017}        & \cm                                & \cm                                             &            & \cm                & \cm          &                &                &                     &                           &                    &                &                       &                    &                      &                    &                    &                &                      &                     &                    &                              &                  &                          &                         &                        \\
                \citet{semanjski_spatial_2017}       & \cm                                &                                                 &            &                    &              &                &                &                     &                           & \cm                &                & \cm                   &                    &                      & \cm                & \cm                & \cm            &                      & \cm                 &                    &                              &                  &                          &                         &                        \\
                \citet{zong2017identifying}\tnote{*} & \cm                                & \cm                                             &            &                    & \cm          & \cm            &                &                     &                           &                    &                &                       &                    &                      &                    &                    &                & \cm                  &                     &                    &                              &                  &                          &                         &                        \\
                \citet{wang2018travel}\tnote{*}      & \cm                                & \cm                                             &            & \cm                & \cm          & \cm            &                &                     &                           &                    &                &                       &                    &                      &                    &                    &                & \cm                  &                     &                    &                              &                  &                          &                         &                        \\
                \citet{dabiri2018inferring}          & \cm                                & \cm                                             & \cm        & \cm                &              &                &                &                     &                           &                    &                &                       &                    &                      &                    &                    &                &                      &                     &                    &                              &                  &                          &                         &                        \\
                \citet{yazdizadeh_ensemble_2020}     & \cm                                & \cm                                             & \cm        & \cm                &              &                &                &                     &                           &                    &                &                       &                    &                      &                    &                    &                &                      &                     &                    &                              &                  &                          &                         &                        \\
                \citet{markos2020unsupervised}       & \cm                                & \cm                                             & \cm        &                    &              &                &                &                     &                           &                    &                &                       &                    &                      &                    &                    &                &                      &                     &                    &                              &                  &                          &                         &                        \\
                \citet{li2020coupled}                & \cm                                & \cm                                             & \cm        & \cm                &              &                &                &                     &                           &                    &                &                       &                    &                      &                    &                    &                &                      &                     &                    &                              &                  &                          &                         &                        \\
                \citet{sadeghian2022stepwise}        & \cm                                & \cm                                             &            & \cm                & \cm          & \cm            &                &                     & \cm                       & \cm                & \cm            & \cm                   &                    &                      & \cm                & \cm                & \cm            &                      & \cm                 & \cm                &                              &                  &                          &                         &                        \\
                \citet{kim2022gps}                   & \cm                                & \cm                                             & \cm        & \cm                &              &                &                &                     &                           &                    &                &                       &                    &                      &                    &                    &                &                      &                     &                    &                              &                  &                          &                         &                        \\
                \citet{roy2022assessing}             & \cm                                &                                                 &            & \cm                & \cm          &                & \cm            &                     &                           &                    &                &                       &                    &                      & \cm                &                    & \cm            & \cm                  &                     &                    &                              &                  & \cm                      & \cm                     & \cm                    \\
                \citet{wu_travel_2022}               & \cm                                & \cm                                             & \cm        & \cm                &              &                &                &                     &                           & \cm                &                & \cm                   &                    &                      &                    &                    & \cm            &                      &                     &                    &                              &                  &                          &                         &                        \\
                \citet{yang_data_driven_2022}        & \cm                                &                                                 &            &                    & \cm          & \cm            &                &                     &                           & \cm                & \cm            & \cm                   &                    &                      &                    &                    & \cm            &                      &                     &                    &                              &                  &                          &                         &                        \\
                \citet{zeng_trajectory_2023}         & \cm                                & \cm                                             &            & \cm                &              & \cm            &                &                     & \cm                       &                    &                &                       &                    &                      &                    &                    & \cm            &                      &                     &                    &                              &                  &                          &                         &                        \\
                \midrule

                Total                                & 25                                 & 17                                              & 6          & 11                 & 10           & 7              & 1              & 4                   & 2                         & 8                  & 7              & 8                     & 1                  & 1                    & 4                  & 3                  & 13             & 6                    & 4                   & 1                  & 1                            & 1                & 1                        & 1                       & 1                      \\
                \bottomrule
            \end{tabular}%
            \begin{tablenotes}
                \item[*] Geospatial context features were only used to detect subway mode by designed rules.
            \end{tablenotes}
        \end{threeparttable}
    \end{adjustbox}
\end{table}

% Motion features 
Overall, motion features have been introduced since the emergence of the problem and are still the most widely used input variables.
Speed and acceleration features can be found in nearly all studies and are considered the most straightforward indicators for distinguishing travel modes~\citep{Tsui2006enhanced, xiao2015travel}.
The bearing rate that reflects the heading change stability and the length of the stage is mainly applied to separate motorized and non-motorized travels~\citep{stenneth2011transportation, xiao2015travel}.
Jerk, which measures the change rate in acceleration, has gained popularity due to its introduction as an additional channel input for deep learning (DL) models~\citep{dabiri2018inferring}.
Other motion features, including duration, altitude, GNSS accuracy, and distance between track points, have been less often employed in recent years.
Operationally, variables observed per track point, such as speed and acceleration, need to be aggregated into a single value to describe a movement trajectory.
Apart from the most often calculated average value, researchers note that a nearly maximum value such as the 85th percentile should be used as an additional indicator~\citep{biljecki2013transportation}, allowing for robustness to noise~\citep{schuessler_processing_2009}.
In addition, a few methods consider more sophisticated statistical indicators (e.g., standard deviation, mode, and skewness) to describe the variable's distribution over the movement trajectory~\citep{xiao_identifying_2017, wu_travel_2022}.

% Geospatial context features
Comparatively, geospatial context data has been used less frequently for travel mode detection.
Based on their represented context information, we divide these features into four categories:
\begin{itemize}
    % network features
    \item Infrastructure networks. These features quantify the trajectory's proximity to networks that allow movements with specific travel modes, typically implemented using different distance measures.
          For example, \citet{stenneth2011transportation} represented the rail network feature with the average Euclidean distance between each track point and its closest rail line. \citet{roy2022assessing} adopted the Hausdorff distance to obtain the furthest distance between track points and the network. Threshold-based methods, which assess the proportion of track points with Euclidean distances closer to the network than a specified threshold, have also been employed in previous studies~\citep{rasmussen2015improved, wu_travel_2022,yang_data_driven_2022}.
    \item Public transport stations. They are the most commonly implemented geospatial context features because of their easy accessibility from open-source map services and their effectiveness in distinguishing public transport from car travel~\citep{chen_evaluating_2010}. Proximity to public transport stations is either considered for the whole trajectory (e.g., by measuring the distance to each track point)~\citep{stenneth2011transportation, roy2022assessing} or for the movement's start and end points (e.g., by only considering the trajectory's endpoints)~\citep{zong2017identifying, wang2018travel}.
          % Public transport timetables and real-time location
    \item Public transport timetables and real-time locations. Besides static geospatial contexts, previous studies integrate information that reflects the actual public transport service situations~\citep{sadeghian2022stepwise, stenneth2011transportation}. However, this dynamic information is only openly accessible in limited areas of the world (e.g., large cities), hindering their application in large-scale travel mode detection studies.
          % Land use and land cover
    \item Land use and land cover (LULC). Relatively few studies consider LULC contexts in the task. \citet{biljecki2013transportation} proposed using the water body feature to identify boat movements, and \citet{roy2022assessing} argued that including LULC features help distinguish non-motorized travel modes. It has been long recognized that the built environment influences an individual's travel mode choice \citep{cheng_applying_2019, tamim_kashifi_predicting_2022}, suggesting a considerable potential to incorporate LULC features for travel mode detection.
\end{itemize}

Although many studies have recognized the importance of geospatial context features~\citep{biljecki2013transportation, semanjski_spatial_2017}, few have attempted to quantify their significance in improving mode detection performance.
Comparative studies have demonstrated that including geospatial context information can significantly increase accuracy, ranging from 18\%~\citep{stenneth2011transportation} to 77~\%~\citep{roy2022assessing}.
However, due to variations in the implemented features, it is challenging to assess and compare the contribution of different context categories to the outcome, hindering the development of an efficient travel mode detection model.

\subsection{Travel mode detection methods}

The widespread use of smartphone GNSS sensors for collecting travel diaries has spurred researchers to develop approaches for automatically detecting travel modes from raw GNSS tracking data.
As a first attempt, rule-based heuristics with human-crafted rule sets to differentiate each travel mode were proposed~\citep{stopher2008search, chen_evaluating_2010, gong2012gps}. These systems match experts' understanding of mode usage but failed to perform satisfactorily when applied to noisy real-world GNSS records.
Therefore, statistical methods such as fuzzy logic~\citep{schuessler_processing_2009, biljecki2013transportation} and Bayesian networks~\citep{xiao2015travel} were applied to account for ambiguity in allocating modes to observed motion and context characteristics.
Later attempts introduced ML for learning classification ``rules'' directly from input features, allowing more flexibility in the decisions and being more robust to real-world noise.
Examples of ML-based approaches include decision trees~\citep{zheng2008learning}, RF~\citep{wang2018travel}, support vector machines~\citep{semanjski_spatial_2017}, and artificial neural networks~\citep{roy2022assessing}.
Among these, RF has become increasingly popular as it can effectively handle high feature dimensionality and multicolinearity~\citep{fernandez_we_2014}. Studies comparing various mode detection methods have consistently found that RF achieved the best performances among classical ML algorithms~\citep{stenneth2011transportation, dabiri2018inferring, sadeghian2022stepwise}.

Thanks to the availability of large-scale datasets, DL models have sparked a new paradigm for travel mode detection.
Examples include convolutional neural network (CNN)~\citep{dabiri2018inferring, yazdizadeh_ensemble_2020} and recurrent neural network (RNN)~\citep{kim2022gps}, which can learn multi-level representations from input features, enabling them to describe highly non-linear relationships.
Additionally, DL models can effectively capture consecutive travel mode choice patterns~\citep{zeng_trajectory_2023}, which is challenging to consider in the standard three-step mode detection framework~\citep{bolbol_inferring_2012}.
However, current DL mode inference models only accept a limited motion feature set as input (e.g., speed, acceleration, jerk, and bearing in~\citet{dabiri2018inferring}), and approaches to include relevant geospatial context information are still to be explored.
The modeling approaches and feature attribution results obtained from this study offer valuable insights that can inspire geospatial context integration into DL models for travel mode detection.

\section{Methodology}\label{sec:method}

\begin{figure}[!htb]
    \centering
    \includegraphics[width=0.8\textwidth]{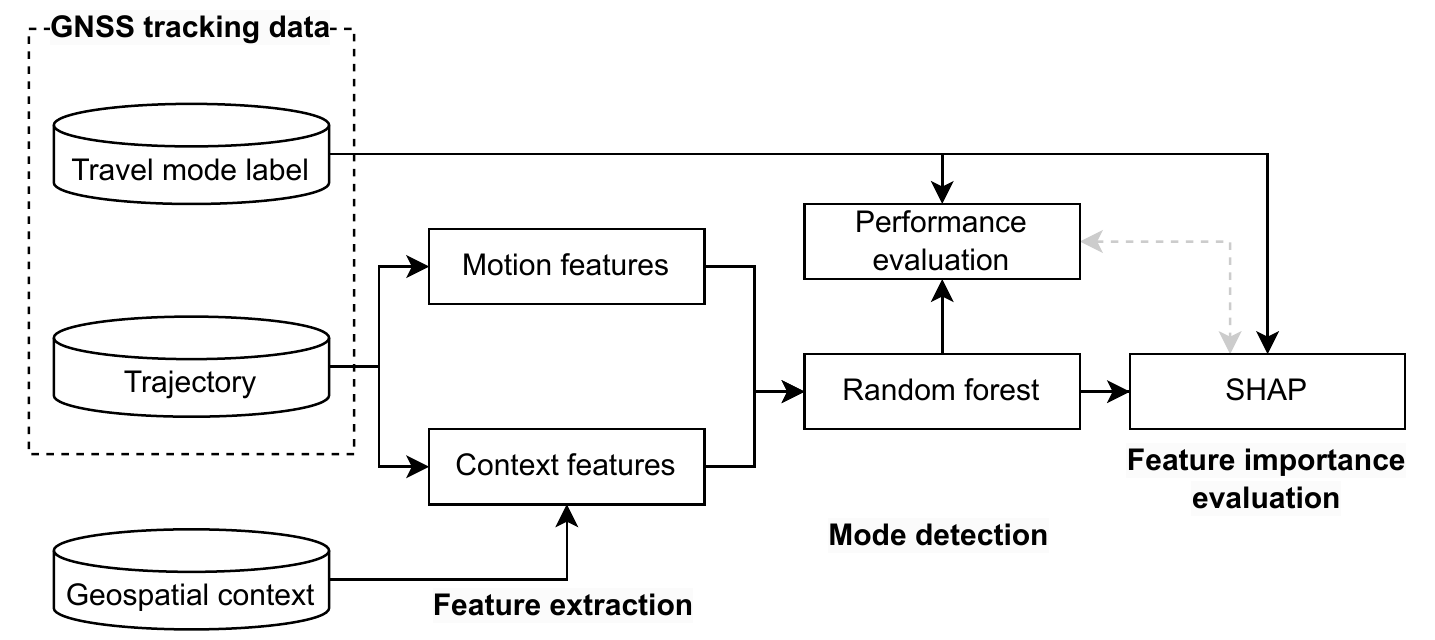}
    \caption{Pipeline for evaluating the contribution of geospatial context information for travel mode detection.}
    \label{fig:overview}
\end{figure}

We present a framework for evaluating the importance of geospatial context information in travel mode detection. The overall pipeline is illustrated in Figure \ref{fig:overview}.
% feature extraction
First, we extract motion and geospatial context features from the GNSS movement trajectory ($\S$\ref{sec:feature}). These features provide a comprehensive characterization of movements performed with different travel modes.
% classifier
Then, we implement an RF classifier to identify the travel mode with the extracted feature set ($\S$\ref{sec:rf}).
Finally, based on the classification outcomes, we evaluate the contribution of the features to the model's prediction using SHAP ($\S$\ref{sec:importance}).
In the following, we provide a more detailed description of each step.

\subsection{Feature Extraction} \label{sec:feature}

As a first step, a typical travel mode identification framework requires extracting meaningful features.
We represent a movement trajectory traveled with a single travel mode by user $u^i$ using $S^i=\langle m, c, g(s) \rangle$, where $m$ represents the employed travel mode, $c$ is the associated geospatial context, and $g(s)$ denotes the time-ordered track points that constitute the movement, i.e., $g(s) = \left (q_k  \right )_{k=1}^{n}$.
A track point $q$ is a tuple of $q = \langle p, t\rangle $, where $p = \langle x, y\rangle $ includes spatial coordinates in a reference system, e.g., latitude and longitude, and $t$ is the time of recording.
Following the notation, we distinguish between motion features, where only track points $g(s)$ are involved, and geospatial context features, where both context information $c$ and track points $g(s)$ are needed.

\begin{table}[htbp!]
    \caption{Description of motion features.}
    \label{tab:motion_features}

    \centering
    \begin{tabular}{@{}lll@{}}
        \toprule
        \textbf{ID} & \textbf{Name}                   & \textbf{Feature} \\ \midrule
        1.1         & Length                          & $D$              \\
        1.2         & Duration                        & $T$              \\
        1.3         & Average speed                   & $\overline{V}$   \\
        1.4         & 85th percentile of speed        & $V^{85^{th}}$    \\
        1.5         & Average acceleration            & $\overline{A}$   \\
        1.6         & 85th percentile of acceleration & $A^{85^{th}}$    \\
        1.7         & Average bearing rate            & $\overline{B}$   \\
        1.8         & 85th percentile of bearing rate & $B^{85^{th}}$    \\ \bottomrule
    \end{tabular}

\end{table}

\paragraph{Motion Feature}
Table~\ref{tab:motion_features} provides an overview of all motion features included in the study. Basic movement metrics such as length, duration, speed, acceleration, and bearing rate are calculated from track points. We obtain the average and the 85th percentile value of these metrics for each trajectory.
More specifically, we calculate the length $\Delta d_k$ and duration $\Delta t_k$ of travel between two consecutive track points $q_k,\,q_{k-1} \in \left (q_k  \right )_{k=1}^{n}$:
\begin{linenomath}
    \begin{equation}
        \Delta {d}_{k}=\|p_{k}-p_{k-1}\|_{2} \qquad \Delta {t}_{k}= t_{k}-t_{k-1}
    \end{equation}
\end{linenomath}
where $\|\cdot\|_{2}$ denotes the Euclidean distance with $p_{k}$ and $p_{k-1}$ represented in a planar coordinate system.
The length $D$ and duration $T$ are obtained through summarizing all these intervals, i.e., $D=\sum_{k=2}^{n} \Delta {d}_{k}$, $T=\sum_{k=2}^{n} \Delta {t}_{k}$. Moreover, the speed $v_k$ and acceleration $a_{k}$ of track point $q_k \in \left (q_k  \right )_{k=1}^{n}$ are obtained as follows:
\begin{linenomath}
    \begin{equation}
        v_k= \frac{\Delta {d}_{k}}{\Delta {t}_{k}} \qquad a_{k}=\frac{v_{k}-v_{k-1}}{\Delta t_{i}}
    \end{equation}
\end{linenomath}

Another essential feature that distinguishes travel modes is the direction change, which is often represented using the bearing rate~\citep{yazdizadeh_ensemble_2020, sadeghian2022stepwise} that measures the absolute difference between the bearings of two sequential track points. The bearing rate $b_k$ of track point $q_k\in \left (q_k  \right )_{k=1}^{n}$ can be calculated as follows:
\begin{linenomath}
    \begin{equation}
        b_k=\left|\beta_{k+1}-\beta_{k}\right|,\quad \text{where}\  \beta_{k}=\arctan (x_{k}-x_{k-1}, y_{k}-y_{k-1})
    \end{equation}
\end{linenomath}

We then calculate the average speed $\overline{V}$, average acceleration $\overline{A}$ and average bearing rate $\overline{B}$ of the trajectory by averaging over all its track points:
\begin{linenomath}
    \begin{equation}
        \overline{V}=\frac{1}{n-1}\sum_{k=2}^{n} {v}_{k} \qquad \overline{A}=\frac{1}{n-2}\sum_{k=3}^{n} {a}_{k} \qquad \overline{B}=\frac{1}{n-2}\sum_{k=2}^{n-1} {b}_{k}
    \end{equation}
\end{linenomath}

The 85th percentile speed $V^{85^{th}}$, acceleration $A^{85^{th}}$ and bearing rate $B^{85^{th}}$ are obtained by ordering the $\left (v_k  \right )_{k=2}^{n}$, $\left (a_k  \right )_{k=3}^{n}$ and $\left (b_k  \right )_{k=2}^{n-1}$ sequences in ascending order and selecting the 85th percentile value, respectively.

\paragraph{Geospatial Context Feature}

\begin{figure}[!t]
    \centering
    \includegraphics[width=1\textwidth]{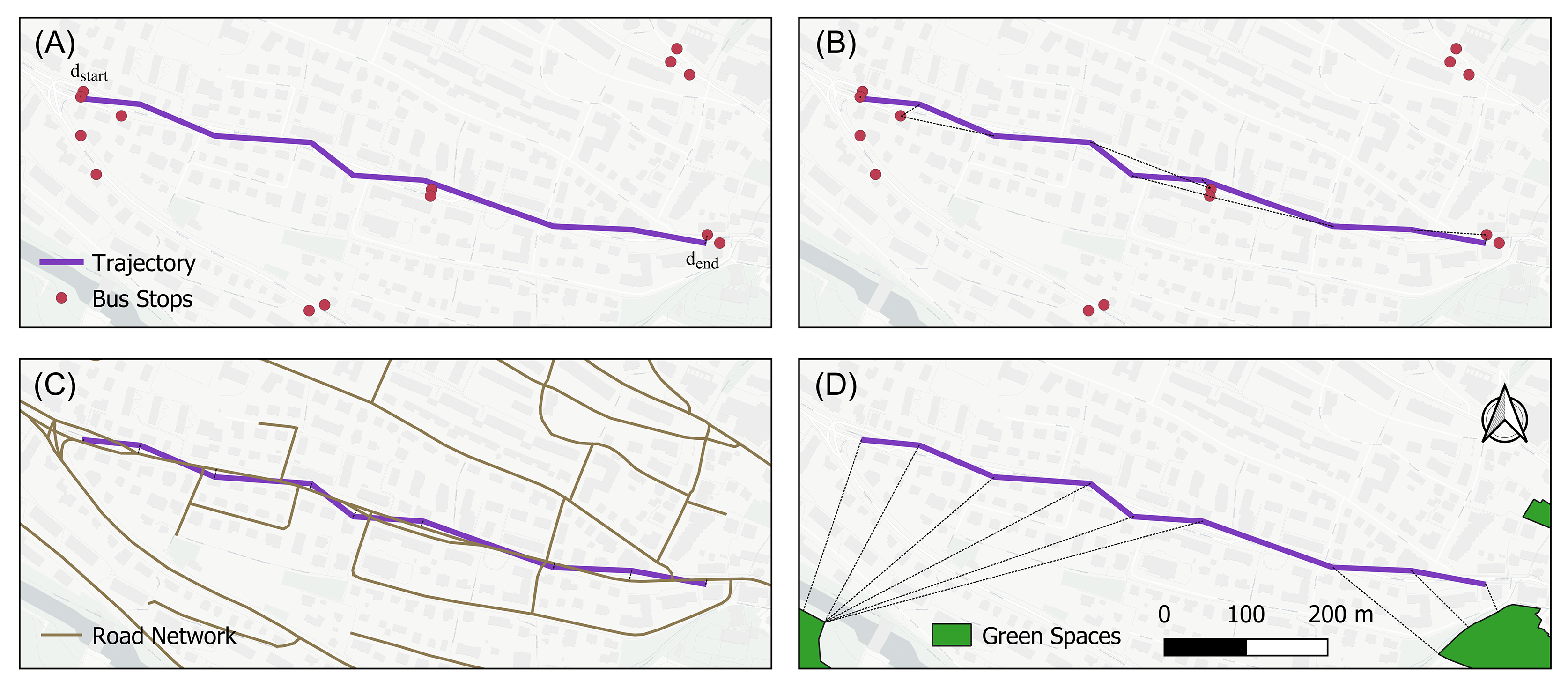}
    \caption{Illustration of geospatial context feature calculation using a sample GNSS trajectory. We showcase the process of constructing features for the trajectory's endpoints with point context (A) and for all trajectory points with point context (B), network context (C), and area context (D). Dotted lines depict the calculated distances. Map data ©OpenStreetMap contributors, ©CARTO.}
    \label{fig:feature_cal}
\end{figure}

Table~\ref{tab:context_features} provides an overview of the geospatial context features considered in the study.
These features are either obtained with the trajectory's endpoints (Table~\ref{tab:context_features} Endpoints) or all points that form the trajectory (Table~\ref{tab:context_features} All points). The latter can be further categorized following the geometric type of the context data, i.e., whether the contexts exist in the form of points (e.g., POI), networks (e.g., road network), or areas (e.g., residential area).
We demonstrate the computation of various geospatial context features using an example trajectory in Figure~\ref{fig:feature_cal}.
We include abundant geospatial context features to exhaustively consider different context modeling approaches in previous studies (see Section~\ref{sec:related}).

\begin{table}[htbp!]
    \caption{Description of geospatial context features.}
    \label{tab:context_features}

    \makebox[\textwidth]{\begin{tabular}{@{}lllll@{}}
            \toprule
            \textbf{Level}               & \textbf{ID} & \textbf{Name}                       & \textbf{Geospatial context} & \textbf{Feature}                  \\ \midrule
            \multirow{6}{*}{Endpoints}   & 2.1/2.2     & Min/max distance to railway station & Railway stations            & $D^{min}_{rail}$/$D^{max}_{rail}$ \\
                                         & 2.3/2.4     & Min/max distance to tram stops      & Tram stops                  & $D^{min}_{tram}$/$D^{max}_{tram}$ \\
                                         & 2.5/2.6     & Min/max distance to bus stops       & Bus stops                   & $D^{min}_{bus}$/$D^{max}_{bus}$   \\
                                         & 2.7/2.8     & Min/max distance to car parkings    & Car parkings                & $D^{min}_{car}$/$D^{max}_{car}$   \\
                                         & 2.9/2.10    & Min/max distance to bike parkings   & Bike parkings               & $D^{min}_{bike}$/$D^{max}_{bike}$ \\
                                         & 2.11/2.12   & Min/max distance to landing stages  & Landing stages for ships    & $D^{min}_{ship}$/$D^{max}_{ship}$ \\ \midrule
            \multirow{12}{*}{All points} & 2.13        & Distance to railway stations        & Railway stations            & $D_{railS}$                       \\
                                         & 2.14        & Distance to tram stops              & Tram stops                  & $D_{tramS}$                       \\
                                         & 2.15        & Distance to bus stops               & Bus stops                   & $D_{busS}$                        \\
                                         & 2.16        & Distance to POIs                    & POIs                        & $D_{POIS}$                        \\ \cmidrule(l){2-5}
                                         & 2.17        & Distance to railway network         & Railway network             & $D_{railN}$                       \\
                                         & 2.18        & Distance to tram network            & Tram network                & $D_{tramN}$                       \\
                                         & 2.19        & Distance to road network            & Road network                & $D_{roadN}$                       \\
                                         & 2.20        & Distance to pedestrian/bike network & Pedestrian/Bike network     & $D_{pedN}$                        \\ \cmidrule(l){2-5}
                                         & 2.21        & Proportion on water                 & Lakes and large rivers      & $P_{water}$                       \\
                                         & 2.22        & Distance to public green spaces     & Public green spaces         & $D_{green}$                       \\
                                         & 2.23        & Distance to residential areas       & Residential areas           & $D_{resident}$                    \\
                                         & 2.24        & Distance to forest areas            & Forest areas                & $D_{forest}$                      \\ \bottomrule
        \end{tabular}
    }
\end{table}

We start by implementing features that only depend on the endpoints of a trajectory.
These features measure the closeness of trajectory endpoints to the geospatial context, assuming that trajectories shall start and end at predefined locations~\citep{stopher2008search, gong2012gps}.
Operationally, we regard the point object set $\mathcal{Q}=\left \{ Q_k \right \}_{k=1}^{m(Q)}$ with $m(Q)$ point objects as the context $c=\mathcal{Q}$, and identify the minimum distance between $\mathcal{Q}$ and the trajectory start point $q_{1}$ as well as the end point $q_{n}$, respectively:

\begin{linenomath}
    \begin{equation}
        {d}_{start} =\min(\left \{ \|q_{1}-Q_{k}\|_{2} \mid Q_k\in \mathcal{Q}\right \}) \qquad
        {d}_{end} =\min(\left \{ \|q_{n}-Q_{k}\|_{2} \mid Q_k\in \mathcal{Q}\right \})
    \end{equation}
\end{linenomath}

We use the minimum and maximum of the two distances to quantify the closeness of a trajectory to the point context:
\begin{linenomath}
    \begin{equation}
        D^{\min } =\min ( {d}_{start}, {d}_{end} ) \qquad
        D^{\max } =\max ( {d}_{start}, {d}_{end} )
    \end{equation}
\end{linenomath}

We construct $\mathcal{Q}$ using context data regarding railway station, tram stop, bus stop, car parking, bicycle parking, and ship landing stage, respectively, which results in 12 features in this category (Figure~\ref{fig:feature_cal}A).

Besides, previous studies have measured the proximity of the entire trajectory to geospatial contexts~\citep{semanjski_spatial_2017}, which accounts for dynamic context interactions during the movement process. Considering the same point object set $\mathcal{Q}$, we now measure the average minimum distance between $\mathcal{Q}$ and every point $q\in \left (q_k  \right )_{k=1}^{n}$ that forms the trajectory:
\begin{linenomath}
    \begin{equation}
        {D_{S}}=\frac{1}{n} \sum_{j=1}^{n} \min(\left \{ \|q_{j}-Q_{k}\|_{2} \mid Q_k\in \mathcal{Q}\right \})
    \end{equation}
\end{linenomath}

We implement features measuring the distance to railway stations $D_{railS}$, tram stops $D_{tramN}$, bus stops $D_{busS}$ and general POIs (e.g., public and catering facilities) $D_{POIS}$ using the respective point context data (Figure~\ref{fig:feature_cal}B).

In addition, movements using specific travel modes are confined by their infrastructures, such as trains and trams that operate on established tracks. This characteristic can be quantified by measuring the distance between the trajectory and infrastructure networks. Here, we consider the infrastructure network $\mathcal{N} = \left \{ L_k \right \}_{k=1}^{m(L)}$ that consists of $m(L)$ line objects as the context $c = \mathcal{N}$. We then calculate the average of the closest distance between every trajectory point $q\in \left (q_k  \right )_{k=1}^{n}$ and the network $\mathcal{N}$:
\begin{linenomath}
    \begin{equation}
        \label{eq:dist}
        D_N = \frac{1}{n} \sum_{j=1}^{n} \min (\left \{f(q_j, L_k)\mid L_k \in \mathcal{N} \right \})
    \end{equation}
\end{linenomath}
where $f(q_j, L_k)$ measures the Euclidean distance between the point $q_j$ and the line $L_k$. As a result, network distances of a trajectory to the railway network $D_{railN}$, to the tram network $D_{tramN}$, to the road network $D_{roadN}$, and to the pedestrian and bike network $D_{pedN}$ are obtained (Figure~\ref{fig:feature_cal}C).

The last feature category relates to built and natural environments that are particularly attractive to or only accessible by specific travel modes. These LULC contexts are typically available in area formats and can be denoted using a set $\mathcal{C}=\left \{ E_k \right \}_{k=1}^{m(E)}$ that contains $m(E)$ non-overlapping area objects. Analogous with Equation~\ref{eq:dist}, we include features describing the distance to residential areas $D_{resident}$, public green spaces $D_{green}$ and forest areas $D_{forest}$ (Figure~\ref{fig:feature_cal}D), as travels close to these LULC contexts were reported to be dominated by active modes and short-distance public transport~\citep{semanjski_spatial_2017, roy2022assessing}.
In addition, we obtain the trajectory's proportion on water $P_{water}$ for describing boat travels:
\begin{linenomath}
    \begin{equation}
        P = \frac{1}{n} \sum_{j=1}^{n} \sum_{k=1}^{m(E)} \mathbbm{1}_{\left [\text{within}(q_j, E_k)\right ]}
    \end{equation}
\end{linenomath}
where $\text{within}(q_j, E_k)$ represents the spatial analysis function for determining whether $q_j$ is within $E_k$, $ \mathbbm{1}_{\left [ \cdot \right ]}$ is the indicator function, and $ \mathbbm{1}_{\left [ C \right ]} = 1$ if $C$ is $True$ and $ \mathbbm{1}_{\left [ C \right ]} = 0$ otherwise.

\subsection{Random Forest Classification}\label{sec:rf}

Given a feature set that describes the trajectory, travel mode detection can be viewed as a classification problem that aims to obtain a (non-linear) mapping $g(\cdot)$ between the feature set $\mathcal{X}$ and the ground-truth travel mode label $c$, i.e., $c = g(\mathcal{X})$. When a new trajectory is observed, the learned mapping $g(\cdot)$ is used to predict a travel mode label $\hat{c}$.
Many supervised ML models have been proposed for learning $g(\cdot)$~\citep{dabiri2018inferring}, among which tree-based models, especially the RF model, are reported to achieve optimum performances~\citep{stenneth2011transportation, yang_data_driven_2022}.

RF is an ensemble method based on the decision tree classifier, with a binary tree structure that partitions the feature space into a set of mutually exclusive regions.
These partitions are performed by searching all possible feature splits and selecting the one that maximizes the Gini impurity gain. For a candidate splitting feature $X_k \in \mathcal{X}$, the Gini impurity index is calculated as:
\begin{linenomath}
    \begin{equation}
        \label{eq:gini}
        \operatorname{Gini}(X_k)=\sum_{i=1}^{C} pr_i \cdot (1-pr_i)
    \end{equation}
\end{linenomath}
where $C$ is the number of categories in $X_k$, and $pr_{i}$ represents the sample proportion of the $i$-th class. The gain for a split is calculated by comparing the Gini impurity in the parent node and the one after performing the split. A decision tree is completely built until pre-specified termination criteria are met, or until all leaves are pure, meaning that only samples from the same category are included~\citep{breiman2017classification}.

While the decision tree is robust to the inclusion of irrelevant features and produces inspectable models, it tends to overfit the training set and has low generalization ability~\citep{hastie_random_2009}.
RF significantly alleviates the overfitting issue by maintaining multiple decision trees, each trained with a different training set, constructed by random sampling from the original training dataset with replacement.
In addition to this sample randomization, often referred to as bootstrap aggregating, RF introduces additional randomness in the feature selection process. Only a random subset of features is considered at each node split, ensuring the diversity of the learned decision trees~\citep{breiman_random_2001}.
During prediction, each constructed decision tree outputs a prediction class label, and the majority voting strategy is used to determine the final classification result.

\subsection{Evaluating feature importance}\label{sec:importance}

With a well-trained RF model, we evaluate the contribution of individual features to the prediction result with SHAP~\citep{shap2017general} and TreeExplainer~\citep{shap2020tree_explainer}. SHAP is a game-theoretic approach to explain the output of ML models using Shapely values~\citep{shapley_value_1952} that fairly distribute players' contributions when they collectively achieve an outcome. The concept can be generalized to ML to quantify the contribution of each feature that collectively delivers the model's output~\citep{strumbelj_explaining_2014}. More formally, the Shapley value $\phi_{X_k}$ of feature $X_k$ is its marginal contribution to the model prediction, averaged over all possible models trained with different feature combinations:
\begin{linenomath}
    \begin{equation}
        \label{eq:shap}
        \phi_{X_k}=\sum_{S \subseteq \mathcal{X} \backslash X_k} \frac{|S|! \cdot \left(|\mathcal{X}|-|S|-1\right) !}{|\mathcal{X}| !}    \left(v\left(S \cup \left\{X_k\right \} \right)-v\left(S\right)\right)
    \end{equation}
\end{linenomath}
where $|\cdot|$ denotes the cardinality of a set and $v(S)$ is the prediction value of a model trained with the feature set $S$.
The exact computation of Shapley values for an arbitrary model has proven to be NP-hard~\citep{matsui_np_2001}, posing computational challenges to their widespread adoption. Facing this challenge, \citet{shap2017general} proposed SHAP to estimate Shapley values, and subsequently, TreeExplainer was presented to exactly compute Shapley value explanations for tree-based models (such as RF) in polynomial time~\citep{shap2020tree_explainer}.

We use the TreeExplainer to obtain SHAP value explanations at the level of individual observations. The overall importance $\phi_{X_k}$ of feature $X_k$ is calculated as the average absolute SHAP values over all considered data samples:
\begin{linenomath}
    \begin{equation}
        \phi_{X_k}=\frac{1}{N}\sum_{i=1}^{N} |\phi_{i, X_k}|
    \end{equation}
\end{linenomath}
where $\phi_{i,X_k}$ is the SHAP value for sample $i$ for feature $X_k$, and $N$ is the considered sample size. A higher absolute SHAP value suggests a stronger influence of the feature on the prediction.
Therefore, we can assess the impact of context features in travel mode detection by analyzing $\phi_{X_k}$ for all implemented motion and geospatial context features.

In addition, we implement feature importance assessment methods employed by previous travel mode detection studies~\citep{yang_data_driven_2022, wu_travel_2022}, including mean decrease in impurity (MDI), permutation importance, and drop column importance, to complement and support the result obtained using SHAP. These methods are based on different principles and are described in detail in Appendix~\ref{app:method}.

\section{Experiment}\label{sec:experiment}

\subsection{GNSS tracking data and preprocessing}

We utilize a large-scale longitudinal GNSS tracking dataset for the case study.
The tracking dataset was recorded within the SBB Green Class (GC) E-Car pilot study conducted by the Swiss Federal Railways (SBB) from November 2016 to December 2017~\citep{martin2019begleitstudie}. The study, involving 139 Switzerland-based participants, aimed to evaluate the effect of a mobility-as-a-service (MaaS) offer on individuals' mobility behavior. The participants were provided with a MaaS bundle and were asked to install a GNSS-tracking application on their smartphones that records their daily movement with a high temporal resolution.
Based on motion measurements such as speed and acceleration obtained from built-in smartphone sensors, the application segments the recorded GNSS traces into \textit{stages} of continuous movements and \textit{staypoints} where users are stationary.
It additionally imputes the travel mode labels for stages (also based on motion measurements), which are later confirmed or corrected by the study participants.
We provide an example that illustrates the attributes of GNSS traces in Appendix~\ref{app:preprocess}.
The GC dataset consists of $\sim$230 million GNSS track points, aggregated into 465,195 stages with travel mode labels (car, e-car, train, bus, tram, bicycle, e-bicycle, walk, airplane, boat, coach), and includes information about individual travel behavior for 52,251 user days.

We implement a series of preprocessing steps to prepare the dataset for travel mode detection following previous research on travel behavior analysis~\citep{hong_conserved_2023} and mode detection~\citep{stopher2008search, dabiri2018inferring}. The detailed steps and the resulting tracking quality can be found in Appendix~\ref{app:preprocess}.
After preprocessing, we obtain 365,307 stages with user-validated mode labels grouped into bicycle, boat, bus, car, train, tram, and walk. The travel mode frequency is shown in Table~\ref{tab:mode_number}, suggesting a highly imbalanced number of class labels: ranging from 367 stages for the class boat to 155,177 for the class walk.

\begin{table}[htbp!]
    \centering
    \caption{Travel mode frequency of stages.}
    \label{tab:mode_number}

    \begin{tabular}{@{}ll@{}}
        \toprule
        \textbf{Mode category} & \textbf{Green Class} \\ \midrule
        Bicycle                & 11,948               \\
        Boat                   & 367                  \\
        Bus                    & 9,436                \\
        Car                    & 130,678              \\
        Train                  & 51,470               \\
        Tram                   & 6,231                \\
        Walk                   & 155,177              \\

        \textbf{Total}         & \textbf{365,307}     \\ \bottomrule
    \end{tabular}
\end{table}

\subsection{Geospatial context data}

The geospatial context data used to calculate context features derives from Open Street Map (OSM)\footnote{\url{http://www.openstreetmap.org}} and Swiss Map Vector 25 (SMV25)\footnote{\url{https://www.swisstopo.admin.ch/en/geodata/maps/smv/smv25.html}}.
OSM is an open-source project that provides users with free and easily accessible digital map resources and is considered the most successful and prevailing volunteered geographic information (VGI) project~\citep{hong_hierarchical_2019}.
We retrieve historical feature layers from early 2017 in Switzerland from OSM to match the time frame of the GC tracking study.
These layers include transport infrastructure, traffic-related POI, general POI, places of worship, road and railway infrastructure, as well as land cover type.
We discuss the representation of geospatial context features when abstracting spatial entities as POIs and its impact on mode detection performance in Appendix~\ref{app:represent}.
We note that the pedestrian/bike network utilized to derive feature 2.20 is constructed by considering all road types permitting walking and biking, along with the specific ``pedestrian'' and ``cycleway'' road types from OSM.
As the water layer of the 2017 dataset does not include main lakes across Swiss borders, this layer is taken from the latest version of OSM (mid-2022).
Besides, we retrieve information regarding the size of a river and whether it is navigable for boats from SMV25.
Table \ref{tab:feature} provides an overview of all layers used as input to extract geospatial context features.

\begin{table}[ht!]
    \caption{The sources, geometry types, and descriptions of the considered geospatial context data.}
    \label{tab:feature}
    \centering
    \makebox[\textwidth]{\begin{tabular}{l l l l}
            \toprule
            Source   & Name         & Type    & Description                                         \\
            \midrule
            OSM 2017 & transport    & Point   & Transport infrastructure (e.g. railway stations)    \\
            OSM 2017 & traffic      & Point   & Traffic-related POI (e.g. car parkings)             \\
            OSM 2017 & pois \& pofw & Point   & General POI \& Places of worship                    \\
            OSM 2017 & roads        & Line    & Road infrastructure (e.g. motorways, tracks, paths) \\
            OSM 2017 & railways     & Line    & Railway infrastructure (e.g. railways, trams)       \\
            OSM 2017 & landuse      & Polygon & Land cover type (e.g. residential areas, forests)   \\
            OSM 2022 & water        & Polygon & Lakes                                               \\
            SMV25    & FGT          & Line    & Rivers                                              \\\bottomrule
        \end{tabular}
    }

\end{table}

\subsection{Model training and evaluation metrics}

Our pipeline is implemented in \textit{Python} using trackintel~\citep{Martin_2023_trackintel}, scikit-learn~\citep{scikit-learn} and SHAP~\citep{shap2017general} libraries.
We randomly split the GC dataset into non-overlapping train and test sets with a ratio of 8:2.
We then perform a grid search using five-fold cross-validation on the training set to determine the optimum hyper-parameters.
Detailed information about the implemented RF model, including strategies for randomizing samples and features, as well as the ranges and the final selected hyper-parameter set, is presented in Appendix~\ref{app:tune}.
We finally retrain the RF model using all the training data and evaluate the model performances on the held-out test set.
SHAP values are obtained from test data samples using path-dependent feature perturbation for the shapely value function (Equation~\ref{eq:shap}), whose results are regarded as ``true to the data'' and reflect natural mechanism in the real world~\citep{chen2020true}.

We use the F1 score to assess the performance of our travel mode detection model.
For each travel mode category, the F1 score is the harmonic mean of precision and recall, which are obtained by constructing the confusion matrix and counting the true positive (TP), false positive (FP), and false negative (FN) samples from the model:
\begin{linenomath}
    \begin{align}
        \text{Precision} & = \frac{\text{TP}}{\text{TP} + \text{FP}}                                               \\
        \text{Recall}    & = \frac{\text{TP}}{\text{TP} + \text{FN}}                                               \\
        \text{F1 score}  & = 2 \cdot \frac{\text{Precision} \cdot \text{Recall}}{\text{Precision} + \text{Recall}}
    \end{align}
\end{linenomath}

We use the average F1 score across classes, which weights the performance for each travel mode fairly without considering its number of instances, thus creating a suitable measure for the class imbalance problem.

\section{Results}\label{sec:result}

\subsection{Feature extraction}\label{sec:res_feature}

\begin{figure}[!htbp]
    \centering
    \includegraphics[width=1.0\textwidth]{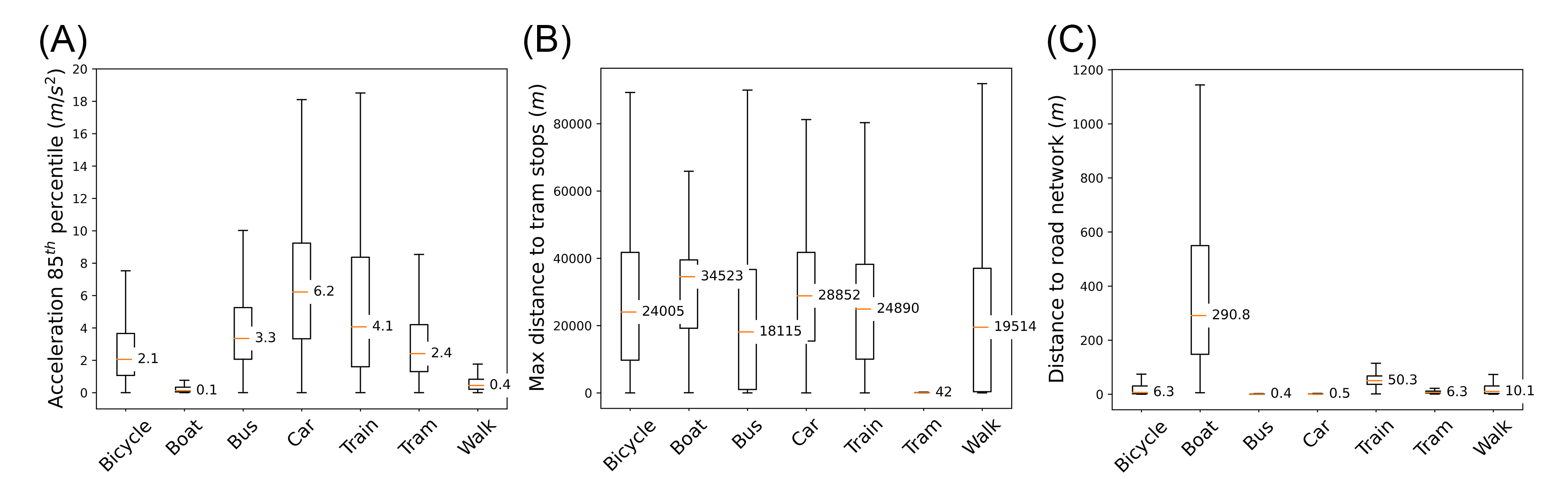}
    \caption{Boxplot showing the feature distribution categorized based on the ground truth travel mode. Three example features 1.6 $A^{85^{th}}$ (A), 2.4 $D^{max}_{tram}$ (B), and 2.19 $D_{roadN}$ (C) are selected to show the effectiveness of the implemented features in separating travel modes. Outlier values are excluded.}
    \label{fig:feature_split}
\end{figure}

We extract 32 features for each movement stage, consisting of 8 motion features and 24 geospatial context features. Figure~\ref{fig:feature_split} shows the distribution of three example features from different categories, highlighting clear distinctions between the considered travel modes.
For instance, \textit{85th percentile of acceleration} distinguishes between low acceleration modes such as boat and walk with high acceleration ones such as car and tram (Figure~\ref{fig:feature_split}A).
Additionally, the stage endpoints for all travel modes except tram have a considerable distance to tram stops (Figure~\ref{fig:feature_split}B).
Finally, \textit{distance to road network} successfully differentiates travel modes that operate on the road network (i.e., bus and car) from those that do not occupy road spaces, such as boat and train (Figure~\ref{fig:feature_split}C).
These examples demonstrate that features characterize movements from different perspectives and facilitate the separation of travel modes.

\begin{figure}[!htb]
    \centering
    \includegraphics[width=0.9\textwidth]{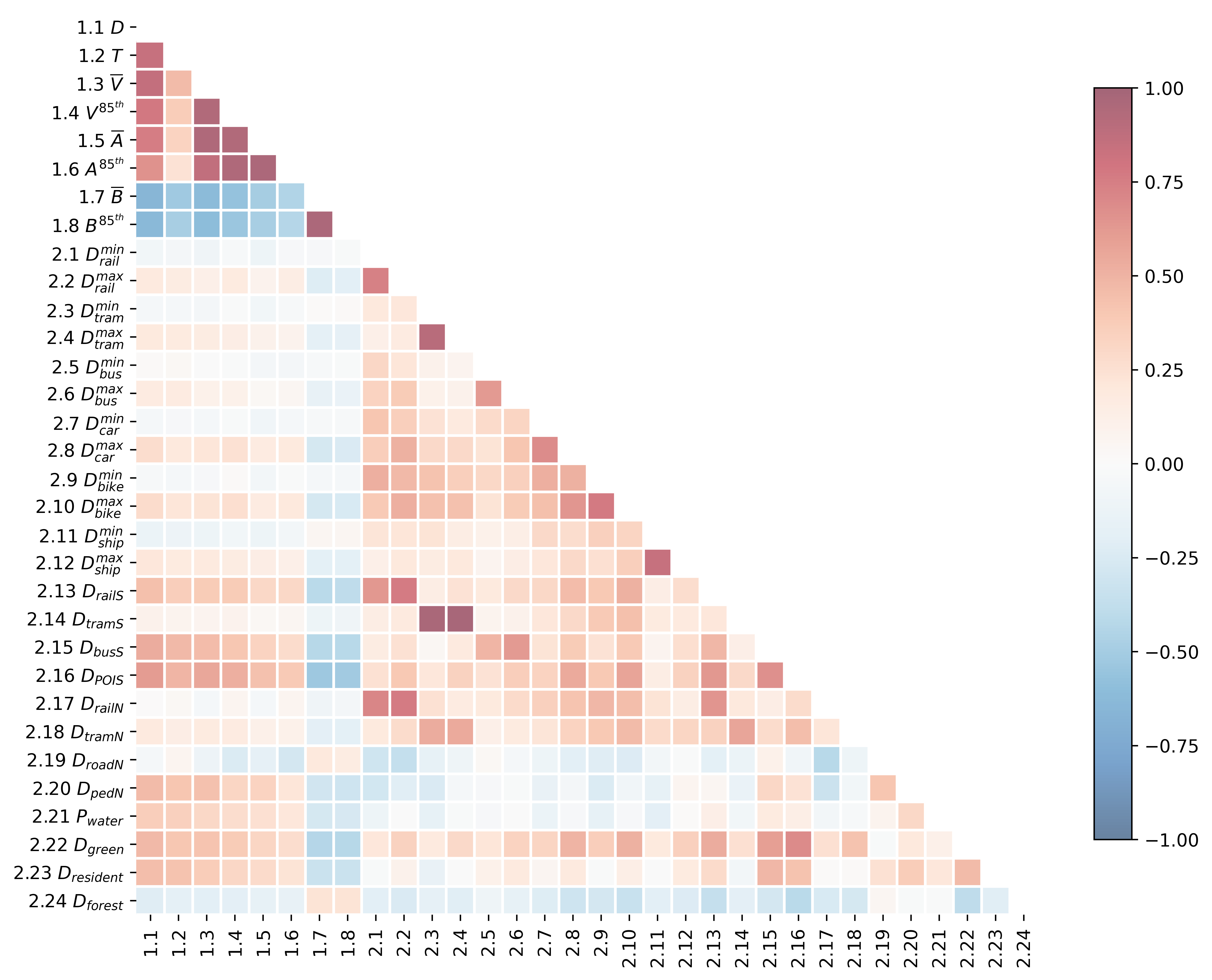}
    \caption{Spearman’s rank correlation coefficient between pairs of features. All pairs have two-tailed $P < 0.05$ except for 2.21 $P_{water}$ and 2.6 $D^{max}_{bus}$ ($P = 0.13$).}
    \label{fig:correlation}
\end{figure}

We conduct a correlation analysis to investigate the relationship between features.
Figure~\ref{fig:correlation} shows the heatmap of Spearman's rank correlation coefficient $\rho$ between pairs of features, which accounts for differences in feature distributions and scales.
The majority of light grid colors representing $\rho$ values close to 0 show that most features are not strongly correlated, indicating that the implemented feature set effectively captures diverse movement characteristics without much redundant information.
However, there are some exceptions. We observe darker grid colors for motion features, showing high positive correlations between length and duration as well as speed and acceleration. We also report high negative correlations between bearing rates and other motion features.
Moreover, we find strong positive correlations between features related to train infrastructure and between features related to tram infrastructure.
Highly correlated features have limited influence on the travel mode detection performance since RF is relatively robust to feature collinearity during training~\citep{fernandez_we_2014}. However, the interpretation of individual feature contributions may be affected depending on the employed feature attribution method~\citep{hastie_random_2009, molnar_interpretable_2020}.

\subsection{Travel mode identification result}

We train an RF model using these features to identify travel modes for movement stages.
The confusion matrix and the precision, recall, and F1 score of travel modes are presented in Table~\ref{fig:matrix}.
We achieve an overall accuracy of 93.0\% and an average F1 score of 83.0\%.
Regarding individual classes, we report reliable detection for the most frequently observed travel modes, namely car, train, and walk, as indicated by their high F1 scores of over 90\%. In addition, the RF model achieved high performance for the tram mode with an F1 score of 96.2\%.
On the other hand, the model experiences difficulty correctly classifying less frequently observed travel modes, such as bicycle, bus, and boat.
Specifically, movement stages recorded as bicycle modes are often misclassified as walk or car modes. This could result from the bicycle mode being formed from both conventional and electric bikes, leading to a wide range of movement behaviors.
We also observe that bus trips are frequently misdetected as car movements, likely due to their similar motion characteristics and shared road infrastructure.
In summary, the overall performance of the mode detection model provides an excellent foundation for estimating the relative importance of each input feature. Yet, we must consider the performance variations between travel modes when interpreting the feature importance result.

\begin{table}[!htb]
    \caption{Confusion matrix and performances for travel mode identification.}
    \label{fig:matrix}

    \centering
    \makebox[\textwidth]{
        \begin{tabular}{@{}lllllllllll@{}}
            \toprule
            \multicolumn{2}{l}{\multirow{2}{*}} & \multicolumn{7}{c}{Predicted class} & \multirow{2}{*}{Sum} & \multirow{2}{*}{Recall (\%)}                                                              \\
            \multicolumn{2}{l}{}                & Bicycle                             & Boat                 & Bus                          & Car    & Train  & Tram   & Walk   &        &               \\ \midrule
            \multirow{7}{*}{Actual class}       & Bicycle                             & 1,191                & 1                            & 45     & 570    & 0      & 0      & 554    & 2,361  & 50.4 \\
                                                & Boat                                & 0                    & 62                           & 0      & 0      & 0      & 0      & 18     & 80     & 77.5 \\
                                                & Bus                                 & 21                   & 0                            & 1,183  & 585    & 1      & 1      & 112    & 1,903  & 62.2 \\
                                                & Car                                 & 334                  & 1                            & 447    & 23,997 & 4      & 5      & 1,333  & 26,121 & 91.9 \\
                                                & Train                               & 2                    & 0                            & 0      & 10     & 10,161 & 8      & 67     & 10,248 & 99.2 \\
                                                & Tram                                & 1                    & 0                            & 0      & 7      & 28     & 1,166  & 38     & 1,240  & 94.0 \\
                                                & Walk                                & 456                  & 12                           & 30     & 448    & 3      & 3      & 30,157 & 31,109 & 96.9 \\ \midrule
            \multicolumn{2}{c}{Sum}             & 2,005                               & 76                   & 1,705                        & 25,617 & 10,197 & 1,183  & 32,279 & 73,062 & -             \\
            \multicolumn{2}{c}{Precision (\%)}  & 59.4                                & 81.6                 & 69.4                         & 93.7   & 99.6   & 98.6   & 93.4   & -      & -             \\ \midrule
            \multicolumn{2}{c}{F1 score (\%)}   & 54.6                                & 79.5                 & 65.6                         & 92.8   & 99.4   & 96.2   & 95.2   & 83.3   & -             \\
            \multicolumn{2}{c}{Accuracy (\%)}   & -                                   & -                    & -                            & -      & -      & -      & -      & 93.0   & -             \\

            \bottomrule
        \end{tabular}
    }
\end{table}

\subsection{Evaluation of Feature Importance}
Based on the trained RF model, we analyze the feature contribution to distinguishing travel modes using various feature importance assessment methods.
This section focuses on feature importance obtained using SHAP values from TreeExplainer, while results obtained from other assessment methods can be found in Appendix~\ref{app:method}.
Figure~\ref{fig:SHAP}A presents the average SHAP values across all test data samples, allowing us to assess the relative importance of features without being influenced by the imbalanced mode distribution.

\begin{figure}[!ht]
    \centering
    \makebox[\textwidth]{
        \includegraphics[width=1.0\textwidth]{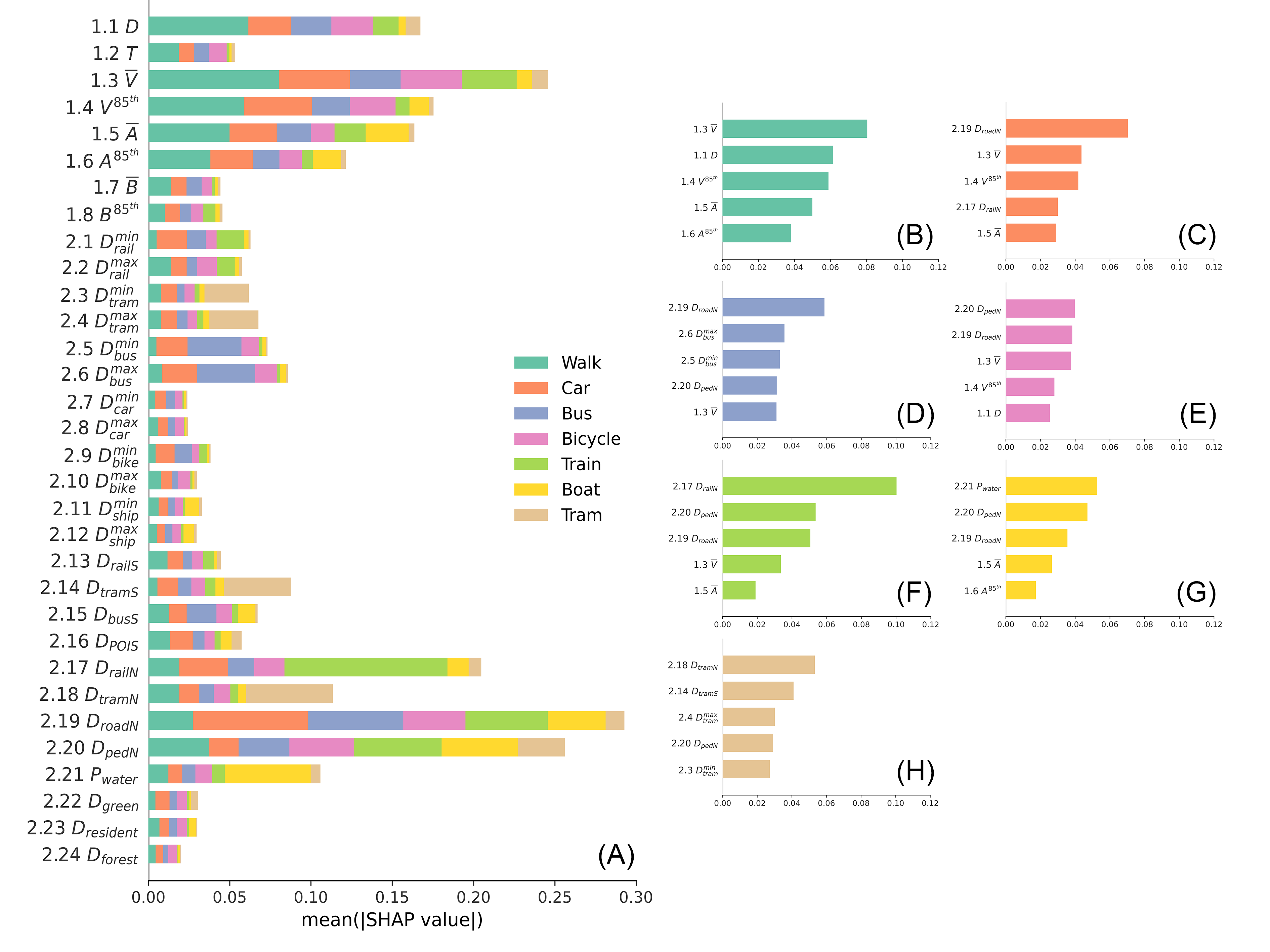}
    }
    \caption{Feature importance evaluated using SHAP. The higher the mean absolute SHAP value, the higher the contribution. We visualize the overall contribution of each feature to the model's output (A) and the top-5 contributing features for identifying walk (B), car (C), bus (D), bicycle (E), train (F), boat (G) and tram (H).}
    \label{fig:SHAP}
\end{figure}

\paragraph{Overall}
We report that \textit{distance to road network} (2.19), \textit{distance to pedestrian and bike network} (2.20), and \textit{average speed} (1.3) contribute the most to the model's output, which belong to geospatial network features and motion features, respectively.
All four network features (2.17 - 2.20) obtained high SHAP values. Within this category, \textit{tram network distance} (2.18) contributes relatively less than the other three features.
Motion features also play a crucial role, as indicated by their high overall importance, particularly distance (1.1), speed-related (1.3 and 1.4), and acceleration-related (1.5 and 1.6) features. However, the SHAP values of duration (1.2) and bearing rate-related (1.7 and 1.8) features are relatively low, suggesting that the detection model does not deem their contained information crucial.
Additionally, geospatial point features moderately contribute to generating the output. Context information regarding public transport, including rail, tram, and bus stops (2.1 - 2.6 and 2.13 - 2.15), is more valuable than the information related to car, bicycle, and ship parking (2.7 - 2.12).
The \textit{distance to POIs} (2.16) is among the most influential features in this category.
Lastly, except for the high contribution of the water body feature (2.21), the other LULC features (2.22 - 2.24) have limited impacts on the detection model, as shown by their lowest SHAP values compared to all other features. We believe that motion and other context features are sufficient in distinguishing travel modes, and the ancillary LULC context information does not provide additional knowledge for the RF model.

\paragraph{Travel mode category}
SHAP values are calculated at the level of individual observations, thus able to reflect detailed feature attribution for each travel mode category.
We presented the five most crucial features for detecting each travel mode in Figure~\ref{fig:SHAP}B-H.
Although these features vary, we observe shared patterns for different travel modes.
The corresponding network feature is always the most important contributing factor for modes restricted to specific infrastructures.
Typical examples include cars, buses, trains, and trams, whose infrastructure network is the most contributing feature, considerably exceeding the importance of the second most crucial one (Figure~\ref{fig:SHAP}C, D, F, and H).
In addition, motion features such as average speed and acceleration are commonly attached to high SHAP values, showing their importance in distinguishing all travel modes.
Their contributions are particularly evident in identifying walking, where the highest SHAP value factors all belong to motion features (Figure~\ref{fig:SHAP}B).
On the contrary, motion features are insufficient in separating bus and tram modes, whose most contributing elements are related to their corresponding geospatial context, such as bus and tram stops (Figure~\ref{fig:SHAP}D and H).
Here, it is evident that identifying bus and tram modes benefit from different context modeling approaches. We observe a high SHAP value of the endpoint-context distance to detect bus trips, whereas the trajectory-context distance contributes more to output tram labels.
Not surprisingly, identifying boat travels benefits the most from the water body feature (Figure~\ref{fig:SHAP}G), which describes the unique characteristics of boats traveling on water.
In summary, this detailed analysis reveals that feature importance is travel mode dependent, and features with low overall importance might be essential for identifying specific travel modes. Besides general motion features, geospatial context features that reveal characteristics of particular modes are also crucial for travel mode detection.

\section{Discussion}\label{sec:discuss}

This study proposes a feature attribution framework for travel mode detection models.
We implement a set of motion and context features to train an RF model, which obtains convincing travel mode identification results.
Note that by considering various mode labels on a large-scale GNSS tracking dataset, the achieved performance is high in view of the recent studies~\citep{yang_data_driven_2022, kim2022gps, wu_travel_2022}. Although the exact performance figures are not directly comparable due to the differences in the employed dataset and preprocessing steps, the result demonstrates the selected features' effectiveness and the RF model's capability. Moreover, the successful detection lays a solid foundation for subsequent feature importance assessment, as feature attribution methods aim to disentangle the decision-making process within the model.
After further validating the model's detection results using official travel surveys~\citep{yang_data_driven_2022, graells_data_2023}, the framework could be applied to automate mode detection in large-scale travel behavior studies that are essential for informing and supporting policy design~\citep{molloy_mobis_2022}.

We use SHAP values to measure the importance of features and compare them with those obtained from other importance assessment methods.
Despite differences in relative importance, all assessment methods regard network features as the most contributing factors, and LULC (except water body) features are of minor significance.
The discrepancies mainly lie in the attribution of motion features, which are most heavily affected by feature correlation.
Permutation and drop column importance both suffer from this issue, underestimating the contributions from motion features. MDI and SHAP reflect the internal decision process of the RF model that is relatively robust to feature colinearity. Consequently, these two methods obtain similar attribution results for motion features.
Nevertheless, MDI utilizes training data to derive its result and cannot reflect the model's generalization ability for unseen data.
Therefore, with a solid foundation from game theory, SHAP is the most suitable and accurate approach to demonstrate the relative importance of geospatial context features.

The SHAP importance value for features helps us understand why the RF detection model may confuse specific travel modes. We show this by comparing each travel mode's most contributing feature set, given in Figure~\ref{fig:SHAP}B-H.
Bicycle movements are detected relying on road and pedestrian networks as well as motion features, which are also the most contributing factors for identifying walking and car trips.
Similarly, the distance to the road network is most influential for both car and bus trips.
The detection model will likely misclassify a travel mode if its distributions in the most contributing features are similar to other modes.
In other words, the lack of distinctive features for bicycle and bus movements limits their prediction performance, leaving room for improving existing and introducing new feature designs.

Although we have systematically evaluated the importance of geospatial context data, there are several points to consider when interpreting the results.
Firstly, it is essential to distinguish the context feature importance obtained from our pipeline from the general significance of the underlying context information. Our study presents only one way to represent context information as features, and feature assessment results may vary with different modeling approaches. However, we note that the features and their representations implemented in this study are carefully selected following a systematic literature review.
Secondly, we exclusively use the average and 85th percentile values to characterize the motion quantities for each trajectory, given their consistent recognition as the primary motion descriptors~\citep{wu_travel_2022, yang_data_driven_2022}. While including comprehensive statistics to depict the distribution of the motion variables might potentially enhance mode detection performance, we do not anticipate a change in the main result and interpretation of the geospatial context feature importance.
Finally, the GC dataset is subject to common GNSS tracking quality issues, such as spatio-temporal gaps and spatial uncertainties in GNSS recordings~\citep{zhao_applying_2021}, which may affect the representativeness of the features and consequently influence the assessment results. Future studies should analyze the robustness of the assessment results to the quality of GNSS tracking data.

\section{Conclusion}\label{sec:conclusion}

Methods proposed for travel mode detection from GNSS tracking data are increasingly powerful, yet little is known regarding the model's underlying working mechanism and how the model outputs a travel mode.
To address this gap, this study introduces an analytical framework for assessing the significance of geospatial context information in travel mode detection models.
Concretely, we review common feature representations from recent work and implement an exhaustive set of features that describe motion characteristics and geospatial context interactions of a moving trajectory.
We analyze the correlations between these features and use them for training an RF model that learns to correctly identify the travel mode.
Using the constructed RF model, we employ feature attribution methods to evaluate the influence of individual features on obtaining the output mode label.
The framework is tested on a longitudinal GNSS tracking dataset containing user-labeled travel modes for trips over 52,000 user days.

The feature attribution results obtained in this study demonstrate that geospatial network features, such as distance to the road network, are more critical than motion features, such as speed and acceleration, when classifying an extensive list of travel modes. This finding highlights the importance of incorporating network features in travel mode detection models, especially given that many existing studies rely heavily on complex motion features without considering the geospatial context.
We also find that features describing relations between movement and geospatial point entities help identify public transport travel with designated start and end stations.
Additionally, our results suggest that the majority of LULC features do not significantly contribute to the task, emphasizing the need for further modeling work to represent the relationship between LULC and movements.
Finally, we identify the most contributing features for detecting each mode, providing insights into the contexts that should be emphasized when aiming to classify specific travel modes accurately.

The proposed travel mode detection framework can be readily applied to movement datasets collected from other parts of the world, thanks to the high-quality and easily-accessible OSM worldwide map service \citep{boeing_using_2022}.
Building upon the findings of this study, we suggest several promising directions to improve the performance of travel mode detection models.
Since this study exclusively focuses on geospatial contexts, the temporal aspects of movements have not been taken into account, despite their considerable influence on individuals' travel mode choices~\citep{schonfelder_urban_2016}. Detection models will likely benefit from including features that differentiate specific travel modes, such as public transport schedules, and features that capture patterns of periodicity and routines in travel behavior.
Furthermore, researchers have extensively examined the trip-chaining effect, including the interdependencies between successive usages of travel modes~\citep{huang_analysis_2021}. However, efforts to integrate this knowledge into detection models remain at an early stage~\citep{zeng_trajectory_2023}.
Lastly, we believe there is potential for refining the extraction of the non-linear relationship between the built environment and travel modes, especially in distinguishing active modes~\citep{cheng_applying_2019}. DL models offer significant promise in this regard. We expect our findings to inspire novel approaches in integrating context for DL models, effectively representing the intricate relationship between movement and geospatial context.
Overall, the study provides valuable guidance for feature selection, effective feature design, and building efficient travel mode detection models.

\bibliography{mybibfile}

\beginsupplement

\newpage
%TC:ignore

\section*{Appendix}

\subsection{Alternative methods for evaluating feature importance}\label{app:method}

\paragraph{Mean decrease in impurity}

The RF model employs the Gini impurity (Equation~\ref{eq:gini}) as the criterion for selecting the splitting feature while constructing its decision tree. This criterion can also be used to evaluate the feature's importance for the classification task. The mean decrease in impurity (MDI) measures the effectiveness of a feature in reducing uncertainty and is calculated by averaging the total Gini impurity reduction of that feature over all decision trees in the RF model. This measure is computationally efficient, but it has two main limitations. First, it overestimates the importance of continuous features and high-cardinality categorical variables. Second, since it is computed on training set statistics, it may not accurately reflect the predictive power of the feature for generalization to the test set.

\begin{figure}[htb]
    \centering
    \includegraphics[width=0.8\textwidth]{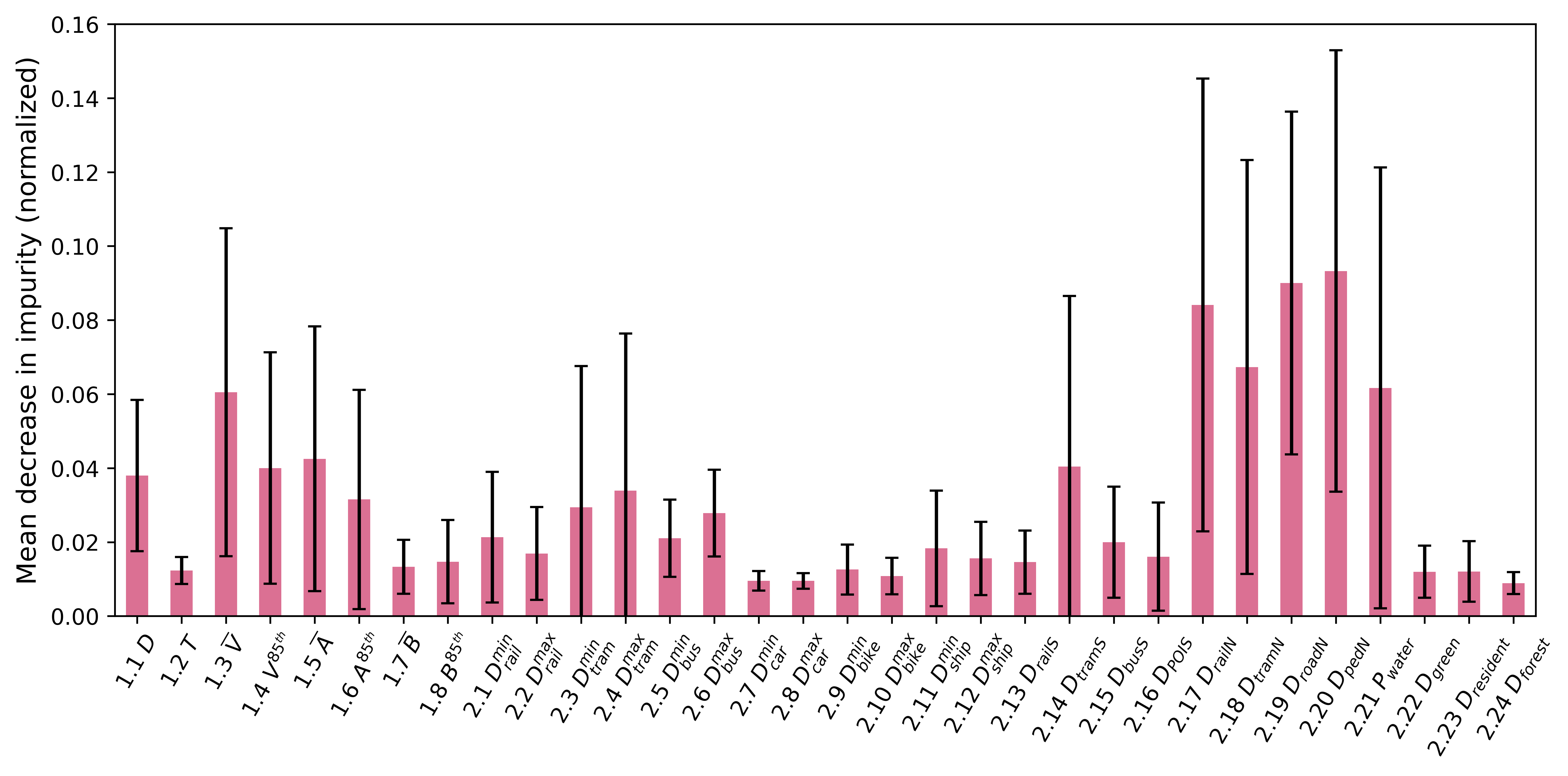}
    \caption{The feature importance obtained from MDI. Error bars represent variability among decision trees.}
    \label{fig:MDI}
\end{figure}

Figure~\ref{fig:MDI} presents the feature importance assessment result, with error bars indicating the variance across different decision trees. The network features are deemed the most important for the RF model, followed by \textit{proportion on water} (2.21) and \textit{average speed} (1.3). The contributions of all other features to the model's prediction are approximately on the same level. The LULC features (2.22 - 2.24) are among the least contributing factors to the model. Moreover, the inter-tree variability is generally very high and positively correlates with the feature importance, likely due to the introduced randomness of feature selection in constructing decision trees.

\paragraph{Permutation Importance}
This measure is computed as the decrease in a model's performance score when a single feature value is randomly shuffled~\citep{breiman_random_2001, molnar_interpretable_2020}. By breaking the feature-target relationship, the drop in the model score indicates how much the model depends on the feature. We compute the permutation importance on the test set without retraining the RF model. However, it is essential to note that this importance score does not reflect the intrinsic predictive value of a feature by itself, but rather how important the feature is for a particular model. For example, features deemed low importance for a poor model could be essential for a good model. Therefore, ensuring a strong predictive power of the model is crucial before using this score. Additionally, permutation importance is biased for highly correlated features. As a well-trained model may still achieve good performance when one of the correlated features is shuffled, the permutation importance measure tends to underestimate the contribution of correlated features.

\begin{figure}[htb]
    \centering
    \includegraphics[width=0.8\textwidth]{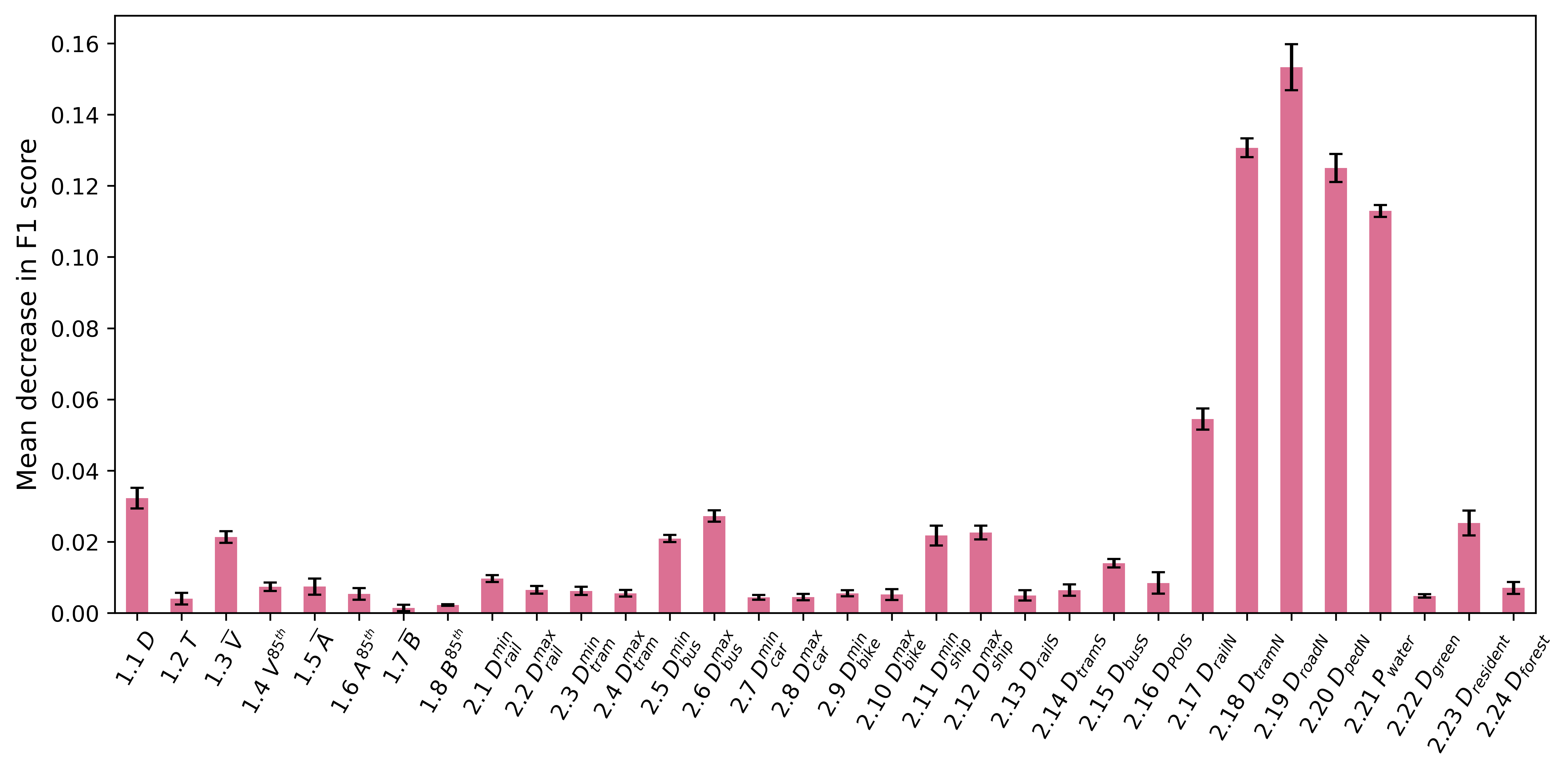}
    \caption{The feature importance evaluated using permutation. The error bars indicate standard deviations calculated from five different permutations.}
    \label{fig:PI}
\end{figure}

Figure~\ref{fig:PI} shows the mean decrease of the F1 score for random shuffling features. Most feature permutations have a limited effect on the model's performance, as indicated by an F1 score decrease of less than 3\%, which might be due to their observed inter-correlations (see Figure~\ref{fig:correlation}). The network features have much higher contributions than all other feature categories, with \textit{distance to road network} (2.19) being the most important among them. Also, including \textit{proportion on water} (2.21) significantly impacts the model's performance. Other decisive contributing factors include motion features (1.1 and 1.3), endpoint features (2.5, 2.6, 2.11, and 2.12), and \textit{distance to residential areas} (2.23). The standard deviation across different permutation runs, indicated by the error bars, is very low, suggesting the permutation importance result is relatively stable.

\paragraph{Drop Column Importance}

The drop column importance method evaluates the contribution of a feature by retraining a model without it. The underlying idea is that training a model without a feature will not significantly affect the performance if the feature is unimportant. However, this method is computationally expensive, requiring retraining the model as many times as the number of features. Additionally, it can underestimate the importance of correlated features. In extreme cases where two features are completely correlated, dropping one will not impact the model, as the remaining feature contains all the necessary information for training, resulting in a zero importance score.

\begin{figure}[htb]
    \centering
    \includegraphics[width=0.8\textwidth]{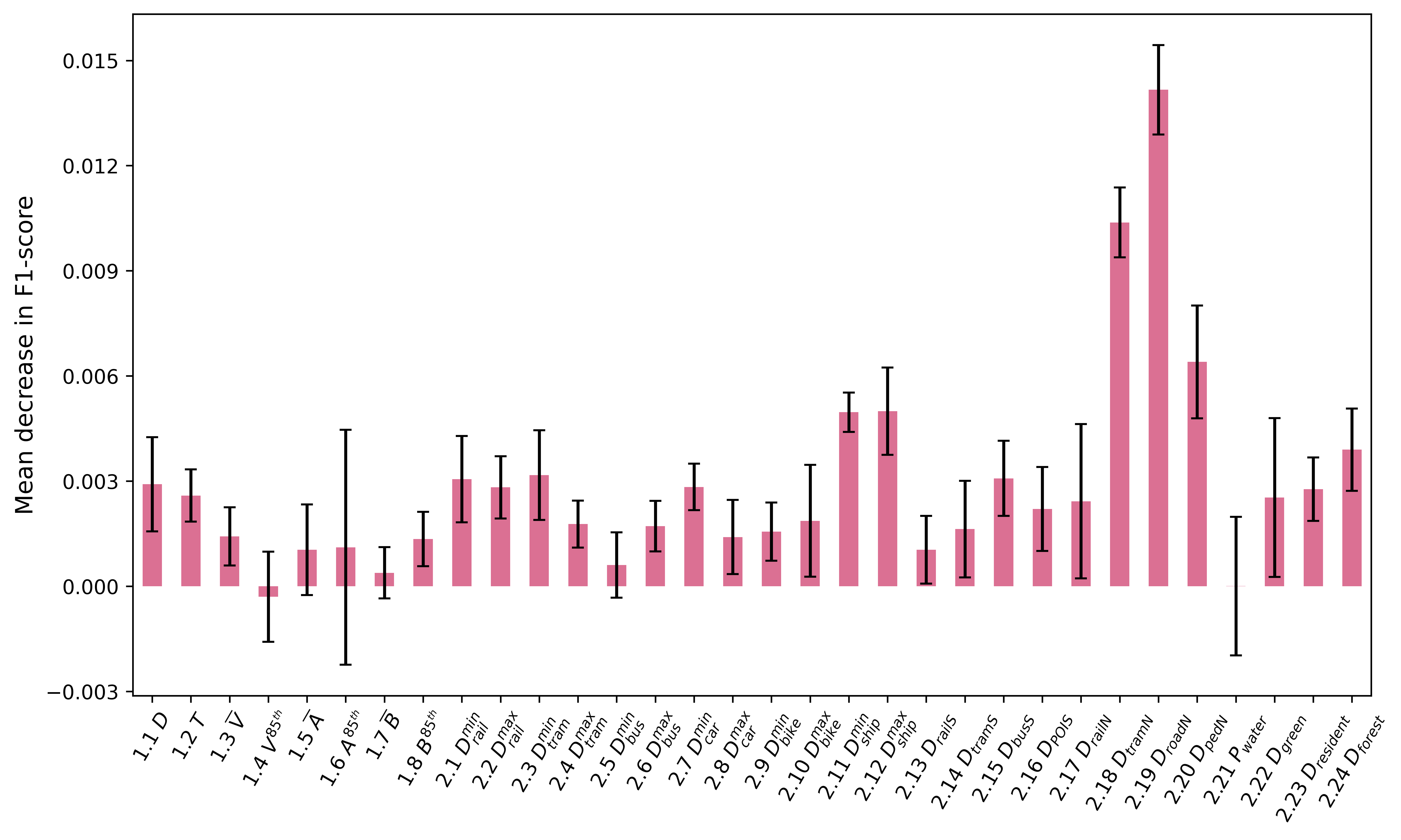}
    \caption{The feature importance evaluated using the drop column method. The error bars represent standard deviations calculated from five different model fits.}
    \label{fig:DCI}
\end{figure}

Drop column importance is measured by the decrease in the F1 score of the test set, averaged across five different model (re-)trainings, as shown in Figure~\ref{fig:DCI}. This method can also produce negative values, which suggests the prediction performance increases when the feature is excluded from the model. Dropping each feature does not significantly influence the model's performance, with most differences in F1 score of less than 1\%. We still observe a relatively high contribution of the network features, especially \textit{distance to road network} (2.19). The distinction between all other features is small. The motion features demonstrated low importance compared to the other assessment methods, most likely due to their strong correlation (see Figure~\ref{fig:correlation}), which diminished their contribution when measured using drop column importance.

\subsection{Preprocessing GNSS tracking dataset}\label{app:preprocess}

Table~\ref{tab:app_GNSS} shows the attributes of the raw GNSS traces collected during the GC study. Each GNSS tracking entry includes details about the user, spatial coordinates, and recording time. In instances where records are affiliated with a stage (as determined by the tracking application; see Section~\ref{sec:experiment}), the application assigns a stage ID and incorporates the user-verified mode label.

\begin{table}[htbp!]
    \caption{Structure and attributes of the collected GNSS trace. The coordinates and timestamps are synthesized to protect the subjects' privacy.}
    \label{tab:app_GNSS}

    \centering
    \begin{tabular}{@{}lllllll@{}}
        \toprule
        \textbf{ID} & \textbf{User ID} & \textbf{Latitude} & \textbf{Longitude} & \textbf{Timestamp}  & \textbf{Stage ID} & \textbf{Mode} \\ \midrule
        1           & CDPXB            & 47.408            & 8.507              & 2016-11-23 07:27:13 & 1                 & walk          \\
        2           & CDPXB            & 47.410            & 8.505              & 2016-11-23 07:27:49 & 1                 & walk          \\
        ...         & ...              & ...               & ...                & ...                 & ...               & ...           \\
        45          & CDPXB            & 47.412            & 8.544              & 2016-12-01 09:43:33 & --                & --            \\
        46          & CDPXB            & 47.413            & 8.542              & 2016-12-01 10:02:28 & 32                & bus           \\
        ...         & ...              & ...               & ...                & ...                 & ...               & ...           \\
        235         & BLWNJ            & 47.498            & 8.719              & 2016-11-24 11:23:06 & 46                & car           \\
        236         & BLWNJ            & 47.497            & 8.717              & 2016-11-24 11:24:01 & 46                & car           \\
        ...         & ...              & ...               & ...                & ...                 & ...               & ...           \\  \bottomrule
    \end{tabular}

\end{table}

We perform data preprocessing to achieve two goals: a) selecting relevant stage data for travel mode detection, and b) filtering incomplete or incorrect data records to ensure high data quality. All processing steps are implemented in \textit{Python} using the open-source \textit{Trackintel} human movement data processing library~\citep{Martin_2023_trackintel}. The steps include:

\begin{itemize}
    \item We pre-filter the dataset to consider high temporal tracking quality participants, using temporal tracking coverage as a measure for the proportion of time the user's whereabouts are recorded. We only include participants with temporal tracking coverage higher than 50\%.
    \item We include only stages recorded within Switzerland to match the geographic extent of the geospatial context data.
    \item We exclude stages labeled with airplane, coach, and ski travel modes. We also merge stages traveled with car and e-car, as well as bicycle and e-bicycle.
    \item We filter out short stages with a total traveled distance of fewer than 50 meters or a total recorded duration of fewer than 60 seconds.
    \item We exclude stages with low tracking qualities, including the ones that consist of less than 4 track points or have average recording intervals of more than one minute. For efficient data storing and processing, we reassigned the timestamp for each track point with a linear interpolation from the start time to the end time of the stage.
    \item We filter stages traveled with unrealistic speeds concerning the reported travel mode. The thresholds are chosen as follows: 20 km/h for walking, 60 km/h for bicycle, 250 km/h for train, 80 km/h for tram, 150 km/h for bus, 50 km/h for boat, and 150 km/h for car.
\end{itemize}

\begin{figure}[htb]
    \centering
    \includegraphics[width=1\textwidth]{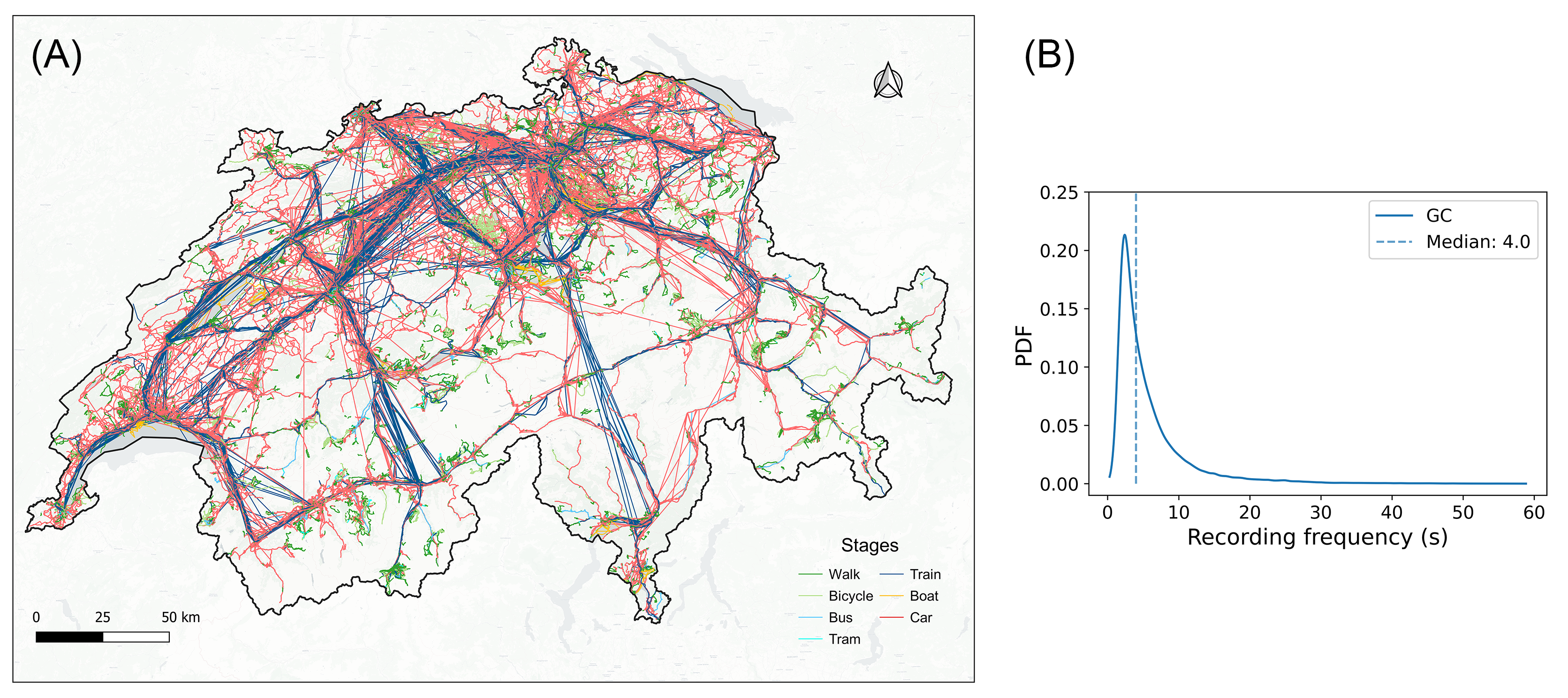}
    \caption{Evaluation of the preprocessed tracking data quality. (A) Spatial distribution of stages conducted using various travel modes. Map data ©OpenStreetMap contributors, ©CARTO. (B) Temporal recording frequency representing the time gap between consecutive GNSS recordings across all stages.}
    \label{fig:freq}
\end{figure}

After preprocessing, we acquire stages featuring high-quality travel mode labels across Switzerland (Figure~\ref{fig:freq}A), comprising GNSS position records sampled at a high frequency (Figure~\ref{fig:freq}B), which enables the precise depiction of motion and geospatial contextual features.

\subsection{Representing spatial entities as POIs}\label{app:represent}

Given their rich attribute information and widespread availability, POIs have been extensively utilized as a data source to represent spatial entities, particularly in addressing urban function-related issues \citep{yao_discovering_2021}.
However, abstracting spatial entities into POIs may introduce spatial errors in the representation depending on the application task, potentially leading to adverse effects on subsequent analyses \citep {psyllidis_points_2022}.
In our study, POIs are employed as proxies for spatial entities. While POIs effectively capture the spatial position of smaller entities, such as bus stops, using them to represent larger entities, such as railway stations and car parking areas, might not accurately convey their spatial relationship with movement stages.
We adopt the movement model defined by \citet{Martin_2023_trackintel}, in which stages are created using all GNSS records between two consecutive stay points where a user remains stationary for a specific duration. This results in the start and end points of stages corresponding to the initial and final movement points, respectively.
Hence, calculating the distance between the track points of a stage and the geospatial context POIs will always yield a non-zero value, even if the track point is located within the spatial entity represented by the POI.
This representation error subsequently affects both the travel mode prediction model and the evaluation of feature importance.

The representation error might not significantly affect the differentiation of travel modes due to: a) the high spatial precision of GNSS recordings and b) the modeling decision to utilize distance as a spatial proximity metric, which is more refined than coarser threshold-based methods.
Nevertheless, we suggest that future research should quantitatively evaluate the influence of spatial entity representations on travel mode detection performance.
Representing these entities as polygon areas rather than point-type POIs and evaluating their topological relationship with movement stages would mitigate this issue, albeit with the trade-off of introducing greater complexity and potentially reducing the efficiency of the modeling framework.

\subsection{Implementing RF model}\label{app:tune}

We employ the scikit-learn Python library \citep{scikit-learn} to implement the randomizing of samples and features for training distinct decision trees within RF. Specifically, we configure the sub-sample set derived from bootstrap aggregating to match the original training set's sample count (controlled by the \textit{max-samples} parameter). The optimal split is ascertained by selecting a random subset of features, with the number of features considered being the square root of the initial feature count (governed by the \textit{max-features} parameter). To ensure consistency across various runs, we employ the \textit{random-state} parameter to manage the introduced randomness in the aforementioned process.

\begin{figure}[!htb]
    \centering
    \includegraphics[width=0.9\textwidth]{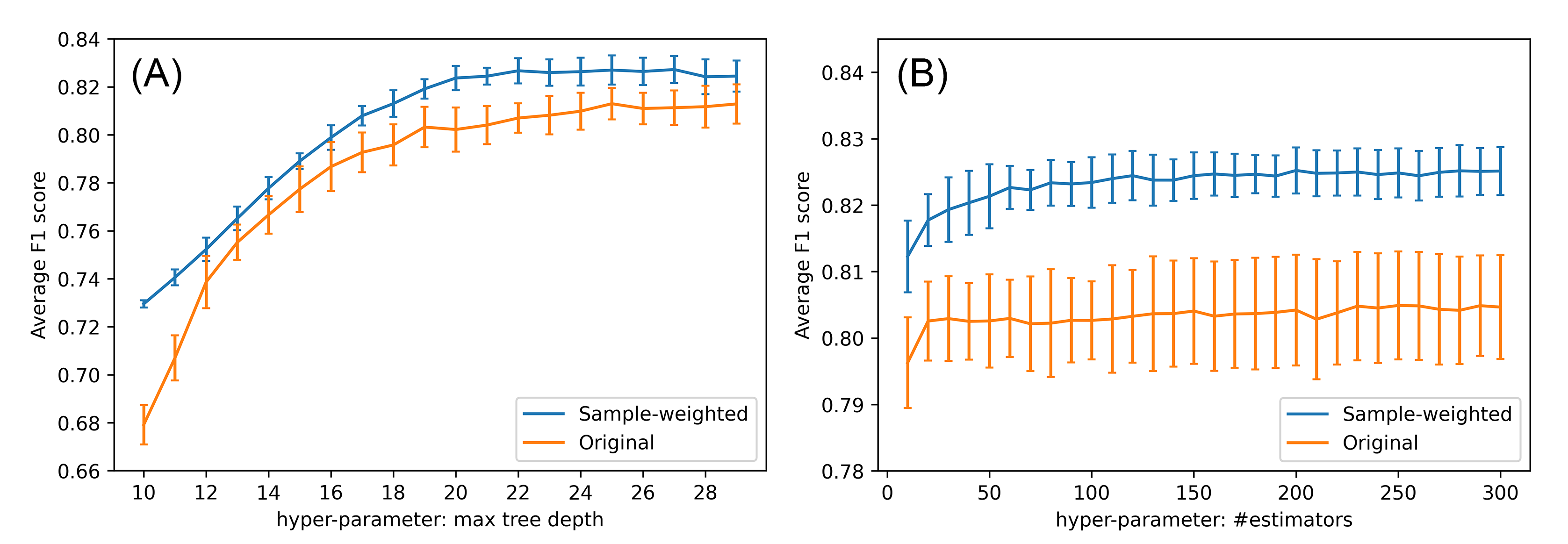}
    \caption{Travel mode prediction performance when altering the hyper-parameters of RF. The error bars represent standard deviations calculated from five-fold cross-validation. We show the performances for sample-weighted and original RF when tuning the maximum tree depth with 150 estimators (A), and tuning the number of estimators with 21 maximum tree depth (B).}
    \label{fig:tuning}
\end{figure}

The hyper-parameter search space includes whether to use sample weighting based on the class frequency (\textit{sample-weighted}), the maximum depth of a tree (\textit{max tree depth}) ranging from 10 to 30 with a step size of 1, and the number of trees (\textit{\#estimators}) ranging from 10 to 300 with a step size of 10. Figure~\ref{fig:tuning} shows the performance variation when increasing max tree depth and \#estimator for both sample-weighted and original RF. We observe the performances initially increase and then gradually stabilize with the increase of max tree depth and \#estimators. The optimum performances are achieved with a sample-weighted RF. We thus select a sample-weighted RF and corresponding parameters where the performance begins to stabilize, i.e., max tree depth=21 and \#estimators=150.

%TC:endignore
\end{document}